\documentclass{jfm}

\usepackage{mathtools}
\usepackage{lmodern}

\usepackage[dvips]{graphicx}
\usepackage{multirow}
\usepackage{url}
\usepackage{setspace}
\usepackage{natbib}
\usepackage[percent]{overpic}
\usepackage{graphics}
\usepackage{graphicx}
\usepackage{amssymb}
\usepackage{amsmath}
\usepackage{cancel}
\usepackage{subfig}
\usepackage{textcomp}
\usepackage{listings}
\usepackage{psfrag}
\usepackage{placeins}
\usepackage{color}
\usepackage{epsfig}
\usepackage{bm}
\usepackage{lscape}
\usepackage[usenames,dvipsnames,table]{xcolor}
\usepackage{tikz}
\usetikzlibrary{positioning,arrows}
\usetikzlibrary{decorations.pathreplacing}
\usepackage{pstricks}
\usepackage[toc,page]{appendix}
\usepackage{upgreek}
\usepackage{caption}
\ifCUPmtlplainloaded \else
  \checkfont{eurm10}
  \iffontfound
    \IfFileExists{upmath.sty}
      {\typeout{^^JFound AMS Euler Roman fonts on the system,
                   using the 'upmath' package.^^J}%
       \usepackage{upmath}}
      {\typeout{^^JFound AMS Euler Roman fonts on the system, but you
                   dont seem to have the}%
       \typeout{'upmath' package installed. JFM.cls can take advantage
                 of these fonts,^^Jif you use 'upmath' package.^^J}%
      }
  \else
  \fi
\fi

\ifCUPmtlplainloaded \else
  \checkfont{msam10}
  \iffontfound
    \IfFileExists{amssymb.sty}
      {\typeout{^^JFound AMS Symbol fonts on the system, using the
                'amssymb' package.^^J}%
       \usepackage{amssymb}%
         \let\leq=\leqslant
         \let\geq=\geqslant
      }{}
  \fi
\fi

\ifCUPmtlplainloaded \else
  \IfFileExists{amsbsy.sty}
    {\typeout{^^JFound the 'amsbsy' package on the system, using it.^^J}%
     \usepackage{amsbsy}}
    {\providecommand\boldsymbol[1]{\mbox{\boldmath $##1$}}}
\fi

\newsavebox{\astrutbox}
\sbox{\astrutbox}{\rule[-5pt]{0pt}{20pt}}

\newcommand\p{\ensuremath{\partial}}

\title[Turbulent drag reduction through oscillating discs]{Turbulent drag reduction \\ through oscillating discs}

\author[D. J. Wise and P. Ricco]%
{Daniel J. Wise%
  \thanks{Email address for correspondence: d.wise@sheffield.ac.uk} \ns
and Pierre Ricco\break
}

\affiliation{Department of Mechanical Engineering, The University of Sheffield,\\
Mappin Street, S1 3JD Sheffield, United Kingdom\\[\affilskip]
}

\pubyear{}
\volume{}
\pagerange{}
\begin{document}

\maketitle



\setstretch{1} 

\begin{abstract}

{\bf This article has been accepted for publication in {\em Journal of Fluid Mechanics}, published by Cambridge University Press.}  The changes of a turbulent channel flow subjected to sinusoidal oscillations of wall flush-mounted rigid discs are studied by means of direct numerical simulations. The Reynolds number is $R_\tau$=$180$, based on the friction velocity of the stationary-wall case and the half channel height. The primary effect of the wall forcing is the sustained reduction of wall-shear stress, which reaches a maximum of 20\%. A parametric study on the disc diameter, maximum tip velocity, and oscillation period is presented, with the aim to identify the optimal parameters which guarantee maximum drag reduction and maximum net energy saving, the latter computed by taking into account the power spent to actuate the discs. This may be positive and reaches 6\%. 

The Rosenblat viscous pump flow, namely the laminar flow induced by sinusoidal in-plane oscillations of an infinite disc beneath a quiescent fluid, is used to predict accurately the power spent for disc motion in the fully-developed turbulent channel flow case and to estimate localized and transient regions over the disc surface subjected to the turbulent regenerative braking effect, for which the wall turbulence exerts work on the discs.

The Fukagata-Iwamoto-Kasagi identity is employed effectively to show that the wall-friction reduction is due to two distinguished effects. One effect is linked to the direct shearing action of the near-wall oscillating disc boundary layer on the wall turbulence, which causes the attenuation of the turbulent Reynolds stresses. The other effect is due the additional disc-flow Reynolds stresses produced by the streamwise-elongated structures which form between discs and modulate slowly in time.

The contribution to drag reduction due to turbulent Reynolds stress attenuation depends on the penetration thickness of the disc-flow boundary layer, while the contribution due to the elongated structures scales linearly with a simple function of the maximum tip velocity and oscillation period for the largest disc diameter tested, a result suggested by the Rosenblat flow solution. 
A brief discussion on the future applicability of the oscillating-disc technique is also presented.

\end{abstract}


\section{Introduction}
Significant effort in the fluid mechanics research community is currently directed towards turbulent drag reduction, motivated by the possibility of huge economic savings in many industrial scenarios. The necessity for improved environmental sustainability has spurred vast academic and industrial interest in the development of novel drag-reduction techniques and in understanding the underlying physical mechanisms.  Although to date there exist many control strategies for drag reduction, notably MEMS-based closed-loop feedback control \citep{kasagi-suzuki-fukagata-2009} and open-loop large-scale wall-forcing control \citep{jung-mangiavacchi-akhavan-1992,berger-etal-2000,quadrio-sibilla-2000}, none have been implemented in industrial systems. Amongst the open-loop active drag reduction methods, for which energy is fed into the system in a pre-determined manner, particular attention has been devoted to those which employ in-plane wall motion. A recent review is found in \cite{quadrio-2011} and a brief 
discussion is presented in the following.
 
\subsection{The oscillating wall} 
 
The direct numerical simulations by \citet{jung-mangiavacchi-akhavan-1992} and the experimental campaign by \citet{laadhari-skandaji-morel-1994} of turbulent wall-bounded flows subjected to sinusoidal spanwise wall oscillations produced a rich vein of work in this area. Their findings first revealed the ability of the actuated wall to suppress the frequency and intensity of near-wall turbulent bursts and to yield a maximum sustained wall friction reduction of about $45\%$. The existence of an optimal oscillation period for fixed maximum wall velocity, $T^+\approx120$ (where $+$ indicates scaling in viscous units with respect to the uncontrolled case) has been widely documented \citep{quadrio-ricco-2004}. 
It was recognized by \citet{choi-xu-sung-2002} that the space-averaged turbulent spanwise flow agrees closely with the laminar solution to the Stokes second problem for oscillation periods smaller or comparable with the optimum one, which led to the use of a scaling parameter for the drag reduction. 
\citet{quadrio-ricco-2004} found a linear relation between this parameter - a measure of the penetration depth and acceleration of the Stokes layer - and the drag reduction, noted to be valid only for $T^+\leq150$. \citet{quadrio-ricco-2004} were also the first to explain the existence of the optimum period by comparing it with the characteristic Lagrangian survival time of the near-wall turbulent structures.
More recently, \citet{ricco-etal-2012} endowed the scaling parameter with a more direct physical meaning, showing it to be proportional to the maximum streamwise vorticity created by the Stokes layer at constant maximum velocity. Through an analysis of the turbulent enstrophy balance, \citet{ricco-etal-2012} were also able to identify the key production term in the turbulent enstrophy equation, which is balanced by the change in turbulent dissipation near the wall. More importantly, by studying the transient evolution from the start-up of the wall motion, they showed that the turbulent kinetic energy and the skin-friction coefficient decrease because of the short-time transient increase of turbulent enstrophy. This is the latest effort aimed at elucidating the drag reduction mechanism, after research works based on the disruption of the near-wall coherent structures \citep{baron-quadrio-1996}, the cyclic inclination of the low-speed streaks \citep{bandyopadhyay-2006}, the weakening of the low-speed streaks \citep{dicicca-etal-2002,iuso-etal-2003}, and simplified models of the turbulence-producing cycle \citep{dhanak-si-1999,moarref-jovanovic-2012,duque-etal-2012}.

\subsection{The wall waves} 

The unsteady oscillating-wall forcing was converted by \citet{viotti-quadrio-luchini-2009} to a steady streamwise-dependent spanwise motion of the wall in the form $\widetilde{W}=W\cos\left(2\pi x/\lambda_x \right)$. Via direct numerical simulations they found an optimal forcing wavelength $\lambda_{opt}^+\approx1250$, which is related to $T_{opt}$, the optimum oscillating-wall period, through $\mathcal{U}_w$, the near-wall convection velocity, as $\lambda_{opt}=\mathcal{U}_wT_{opt}$. \citet{skote-2013} employed Viotti {\em et al.}'s forcing to alter a free-stream turbulent boundary layer and found good agreement between the analytic solution to the spatial Stokes layer flow and the time-averaged spanwise flow. \citet{skote-2013} also showed that the damping of the turbulent Reynolds stresses depends on the penetration depth of the spatial Stokes layer.

The oscillating-wall and the steady-wave techniques were generalized by \citet{quadrio-ricco-viotti-2009} by considering wall turbulence forced by wall waves of spanwise velocity of the form $\widetilde{W}=W\cos\left[2\pi (x/\lambda_x - t/T)\right]$. A maximum drag reduction of 47\% and a maximum net energy saving of 26\% were computed. For wall waves travelling at a phase speed comparable with the near-wall turbulent convection velocity, drag increase was also found.

Despite the widespread interest in turbulent drag reduction by active wall forcing, the implementation of these techniques in industrial settings appears to be an insurmountable challenge. Progress is nonetheless being made to improve this scenario. Prominent amongst recent efforts is the experimental work by \citet{gouder-potter-morrison-2013} on in-plane forcing of wall turbulence through a flexible wall made of electroactive polymers. The main reasons which render the technological applications of active techniques an involved engineering task are \textit{i)} the extremely small typical time scale of the wall forcing (the optimal period for the oscillating-wall technique translates to a frequency of 15,000Hz in commercial aircraft flight conditions), and \textit{ii)} the requirement of large portion of the surface to be in uniform motion. Therefore, drag reduction methods which operate on a large time scale and rely on finite-size wall actuation are preferable in view of future applications.

\subsection{The rotating discs}

The novel actuation strategy based on flush-mounted discs rotating upon detection of the bursting process, first proposed by \citet{keefe-1998}, undoubtedly belongs to a group of interesting control methods which employ finite-size actuators. However, Keefe did not follow up on his innovative idea and neither experimental nor numerical results appeared in the subsequent 15 years. \citet{ricco-hahn-2013} (denoted by RH13 hereafter) showed revived interest in this flow and investigated an open-loop variant of Keefe's technique whereby the discs rotate with a pre-determined constant angular velocity. A numerical parametric investigation on $D$, the disc diameter, and $W$, the disc tip velocity, yielded maximum values for drag reduction and net power saved of 23\% and 10\%, respectively. RH13 also showed that drag increase occurs for small diameter and small rotational periods, that the disc-flow boundary layer must be thicker than a threshold to obtain drag reduction, and that the power spent to activate the 
discs can be calculated accurately through the von K\'arm\'an laminar viscous pump solution \citep{panton-1995} under specified conditions. The Fukagata-Iwamoto-Kasagi (FIK) identity \citep{fukagata-iwamoto-kasagi-2002} was modified for the disc flow to show that the near-wall streamwise-elongated jets appearing between discs provide a favourable contribution to drag reduction. Promisingly, the optimal spatial and temporal scales were $\mathcal{L}^+=\mathcal{O}(1000)$ and $\mathcal{T}^+=\mathcal{O}(500)$. This is a significant result when these scales are compared with those of other localized actuation strategies, such as the feedback control based on wall transpiration \citep{yoshino-suzuki-kasagi-2008}, which are thought to operate optimally at spatio-temporal scales $\mathcal{L}^+=\mathcal{O}(30)$ and $\mathcal{T}^+=\mathcal{O}(100)$. It is our hope that the results of RH13 will therefore offer fertile ground for new avenues of future research on active turbulent drag reduction.

\subsection{Objectives and structure of the paper}

Prompted by RH13's recent results, the objective of the present work is to study a variant of RH13's disc technique by introducing  sinusoidal oscillations, i.e. the disc tip moves according to $\widetilde{W}=W\cos\left(2\pi t/T\right)$. The effect of the additional parameter $T$, the oscillation period, on a turbulent channel flow is investigated through direct numerical simulations, with specific focus on the skin-friction drag reduction and the global power budget, computed by taking into account the power spent to activate the discs. The laminar solution for the flow over an oscillating disc proves useful to estimate the power spent to activate the discs, to predict the occurrence of regenerative braking effect, and to define scaling parameters for drag reduction. An analogy is also drawn to the oscillating wall technique to discuss the drag reduction mechanism at work in the oscillating-disc flow.

The numerical procedures, flow field decompositions and performance quantities are described in \S\ref{sec:numerics}. The solution of the laminar flow is presented in \S\ref{sec:laminar}, where it is used to compute the power spent to move the discs and to predict the regenerative braking effect. The turbulent flow results are presented in \S\ref{sec:turbulent}. The dependence of drag reduction on the disc parameters is discussed in \S\ref{sec:turbulent-time} and \S\ref{sec:turbulent-dependence}. In section \S\ref{sec:turbulent-FIK} the FIK identity is modified to account for the disc flow effects, while \S\ref{sec:turbulent-discflow} presents visualisations and statistics of the disc flow. Section \S\ref{sec:turbulent-pspent} includes a comparison between the turbulent power spent and the corresponding laminar prediction. A discussion on the drag reduction physics and scaling is found in \S\ref{sec:turbulent-scaling-s}. Finally, section \S\ref{sec:outlook} presents an evaluation of the applicability of the 
technique to flows of technological interest, provides a guidance for future experimental studies, and offers a comparison with other drag reduction techniques, with particular focus on the typical length and time scales.

\section{Flow definition and numerical procedures}
\label{sec:numerics}

\subsection{Numerical solver, geometry and scaling}
The simulated pressure-driven turbulent channel flow at constant mass flow rate is confined between two infinite parallel flat walls separated by a distance $L_y^*=2h^*$, where the symbol $^*$ henceforth denotes a dimensional quantity.  The streamwise pressure gradient is indicated by $\Pi^*$. The direct numerical simulation (DNS) code solves the incompressible Navier-Stokes equations in the channel flow geometry using Fourier series expansions along the streamwise ($\tilde x^*$) and spanwise ($\tilde z^*$) directions, and Chebyshev polynomials along the wall-normal direction $y^*$. The time-stepping scheme is based on a third-order semi-implicit backward differentiation scheme (SBDF3), treating the nonlinear terms explicitly and the linear terms implicitly. The discretized equations are solved using the Kleiser-Schumann algorithm \citep{kleiser-schumann-1980}, outlined in \citet{canuto-etal-2007}.  
Dealiasing is performed at each time step by setting to zero the upper third of the Fourier coefficients along the streamwise and spanwise directions. The simulations were carried out using an OpenMP parallel implementation of the code on the N8 HPC Polaris cluster. The code was also used by RH13 and it is a developed version of the original open-source code available on the Internet \citep{gibson-2006}. 

Lengths are scaled with $h^*$ and velocities are scaled with $U_p^*$, the centreline velocity of the laminar Poiseuille flow at the same mass flow rate. The time is scaled by $h^*/U_p^*$ and the pressure by $\rho ^*U_p^{*^2}$, where $\rho^*$ is the density of the fluid. The Reynolds number is $R_p=U_p^* h^*/\nu^*=4200$, where $\nu^*$ is the kinematic viscosity of the fluid. The friction Reynolds number is $Re_\tau=u_\tau^* h^*/\nu^*=180$, where $u_\tau^*=\sqrt{\tau_w^*/\rho^*}$ is the friction velocity in the stationary wall case, and $\tau_w^*$ is the space- and time-averaged wall-shear stress. Quantities non-dimensionalized using outer units are not marked by any symbol. Unless otherwise stated, the superscript $+$ indicates scaling by native viscous units, a terminology first defined by \citet{trujillo-bogard-ball-1997}, based on $u_\tau^*$ of the case under investigation. 
\begin{figure}
  \centering
  \includegraphics[width=0.9\textwidth]{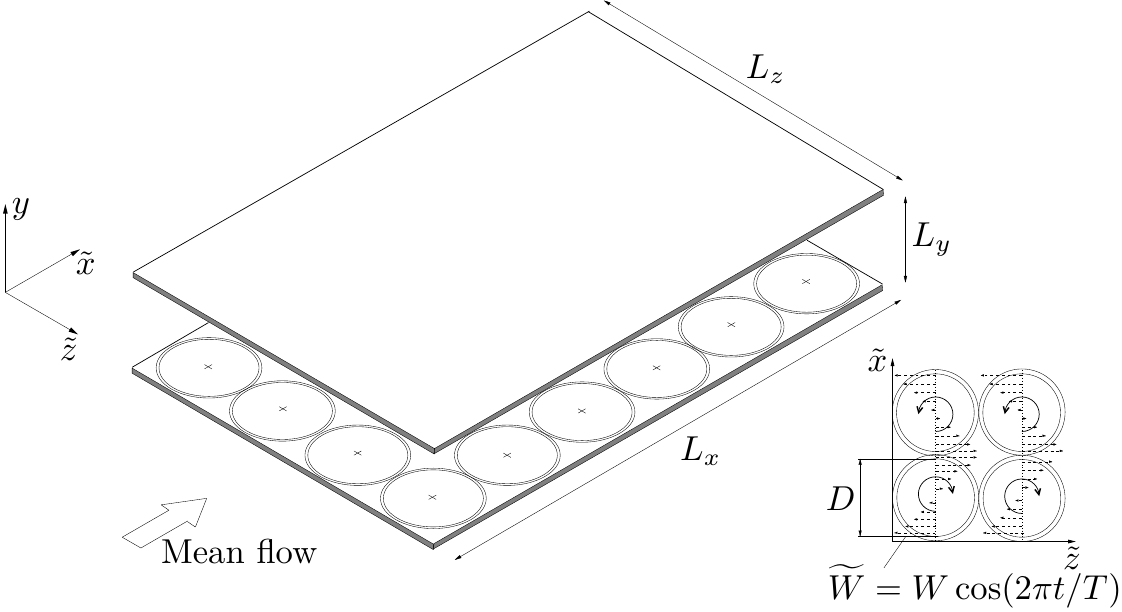}
  \caption{Schematic of the flow domain showing the location and sense of rotation of the discs when $\widetilde{W}=W$.}
  \label{discs-channel}
\end{figure}

The channel walls are covered by flush-mounted rigid discs, as shown schematically in figure \ref{discs-channel}. The discs have diameter $D$ and oscillate in time as the disc tip velocity is
\begin{equation} 
\label{eq:wall-bc}
\widetilde{W}=W\cos \left(\frac{2\pi t}{T}\right). 
\end{equation}
Neighbouring discs in the streamwise direction have opposing sense of rotation, whilst neighbouring discs in the spanwise direction have the same sense of rotation. A parametric study was undertaken on $D$, $W$ and $T$, with the parameter range selected in order to focus on the portion of $D$, $W$ parameter space studied by RH13 which leads to high drag reduction. The region of drag increase found by RH13 was not considered. For disc diameters $D=1.78,3.38$, a computational box size of dimensions $L_x=6.79\pi$ and $L_z=2.26\pi$ was utilized, where $L_x$ and $L_z$ are the box lengths along the streamwise and spanwise directions, respectively. For $D=5.07$, $L_x=6.8\pi$ and $L_z=3.4\pi$, and for $D=6.76$, $L_x=9.05\pi$ and $L_z=2.26\pi$. The grid sizes were $\Delta x^+=10$, $\Delta z^+=5$ in all cases, and the time step was within the range $0.008\leq \Delta t^+\leq0.08$ (scaled in reference outer units). The initial transient period during which the flow adjusts to the new oscillating-disc regime was 
discarded following the procedure outlined in \citet{quadrio-ricco-2004}. Flow fields were saved over an integer number of periods at intervals of $T/8$. After the transient was discarded, the total integration time was  $t^+$$=$6000 for $T^+=100$, $t^+$$=$7500 for $T^+=250, 500$, $t^+$$=$15000 for $T^+=1000$ and $t^+$$=$30000.

\subsection{Model of disc annular gap}

\begin{figure}
  \centering
  \includegraphics[width=0.6\textwidth]{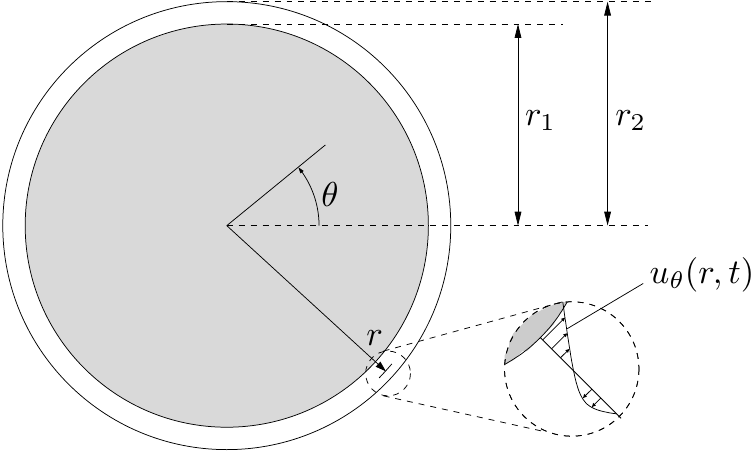}
  \caption{Schematic of disc and annular gap flow.}
  \label{gap-geom}
\end{figure}

To simulate the disc flow as realistically as possible, a thin annular region of width $c$ was simulated around each disc, as shown in figure \ref{gap-geom}. As explained in RH13, there are two reasons for this choice. The clearance flow between each disc and the stationary portion of the wall is simulated to mimic as closely as possible an experimental disc flow set up where such gap would inevitably be present. Secondly, the velocity profile between the disc tip and stationary wall does not present discontinuities. This serves to suppress strongly the Gibbs-type artificial oscillations that would occur if the velocity were not continuous. Ideally, the gap flow would be more realistically simulated by treating the turbulent channel flow and gap flow as coupled systems, but this lies outside the scope of the present study.  

\begin{figure}
  \centering
  \includegraphics[width=\textwidth]{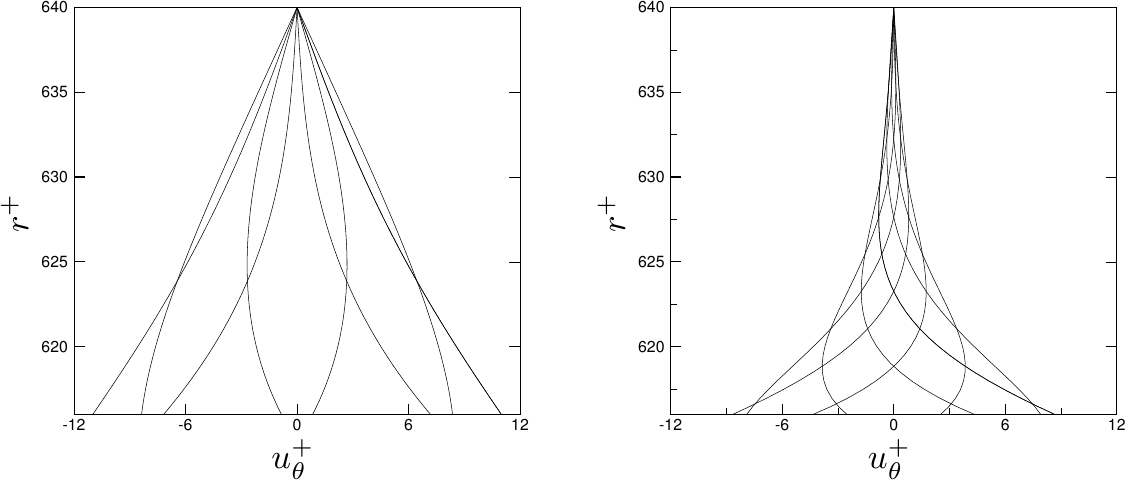}
  \caption{Velocity profiles within the annular gap over a half period of the oscillation, computed through \eqref{utheta-gap}. Left: $D=7.1$, $W=0.51$, $T=130$, $N_w=2.03$. Right: $D=7.1$, $W=0.51$, $T=13$, $N_w=6.42$.}
  \label{fig:gap-profile-oscill}
\end{figure}

As a first approximation, the gap velocity profile is assumed to be symmetric about the disc axis and to change linearly from a maximum velocity at the disc tip to zero at the outer edge of the gap.  The tangential velocity $u_\theta$ in this region is a function only of $r$, the radial displacement from the centre of the disc, and time, $t$. The disc velocity profile is 
\begin{equation*}
  u_\theta(r,t) = \left\{ 
  \begin{array}{l l}
    2Wr\cos(2\pi t/T)/D,
    & 
    \quad 
    r \leq r_1\text{,} \\
    W(c-r+D/2)
    \cos(2\pi t/T)/c,
    & 
    \quad r_1 \leq r \leq r_2\text{,}\\
  \end{array} \right.
\end{equation*}
where $r_1=D/2$ and $r_2=D/2 +c$. As a more advanced approximation, the clearance flow is modelled as a thin layer of fluid confined between concentric cylinders. Similarly to the laminar flow between moving flat plates, the flow contained within this annular gap is described by the Womersley number, $N_w=c^*\sqrt{2\pi/(\nu^*T^*)}$ \citep{pozrikidis-2009}. When $N_w\ll1$, the linear velocity profile accurately describes the flow. However, for $N_w=\mathcal{O}(1)$ the oscillating flow surrounding each disc is confined to a boundary layer which is attached to the oscillating disc and is much thinner than $c$. The bulk of the annular gap is quasi-stationary. In our simulations the minimum $N_w=0.51$ occurs for the case with the thinnest gap and the largest oscillation period, i.e. for $D=1.78$, $T=130$. The maximum $N_w=6.42$ occurs for $D=7.1$, $T=13$. Therefore, it is a sensible choice to simulate the gap via the oscillating layer as $N_w$ attains finite values. 
Following the analysis of \citet{carmi-tustaniwskyj-1981}, the $u_\theta(r,t)$ velocity profile in the gap is described by the azimuthal momentum equation,
\begin{equation}
\frac{\p u_\theta}{\p t} = \frac{1}{R_p}\left(\frac{\p^2u_\theta}{\p r^2} + \frac{1}{r}\frac{\p u_\theta}{\p r} - \frac{u_\theta}{r^2}\right)\text{.}
\label{gap-NS-inner}
\end{equation}
Assuming a solution to \eqref{gap-NS-inner} of the form $u_\theta=\mathbb{R}\left[ \mathring{u}_\theta(r)e^{i2\pi \hat{t}/T} \right]$, where $\mathbb{R}$ denotes the real part and $\hat{t}$ is the rescaled time, $\hat{t}=t/R_p$, the following ordinary differential equation of the Bessel type is obtained
\begin{equation}
\mathring{u}''_\theta + \frac{\mathring{u}'_\theta}{r} - \left(\frac{2\pi i}{T}+\frac{1}{r^2} \right)\mathring{u}_\theta=0,
\label{gap-NS-bess}
\end{equation}
where the prime denotes differentiation with respect to $r$. Equation \eqref{gap-NS-bess} is subject to $\mathring{u}_\theta(r_1)= W$, $\mathring{u}_\theta(r_2)= 0$. The velocity in the annular gap is 
\begin{equation}
u_\theta(r,\hat{t})
= W \cdot\mathbb{R}\left[
\frac{\mathcal{K}(\xi r_2)\mathcal{I}(\xi r)-\mathcal{I}(\xi r_2)\mathcal{K}(\xi r)}{\mathcal{I}(\xi r_1)\mathcal{K}(\xi r_2)-\mathcal{I}(\xi r_2)\mathcal{K}(\xi r_1)}  e^{i2\pi \hat{t}/T} 
\right],
\label{utheta-gap}
\end{equation}
where $\mathcal{I}(\cdot)$ and $\mathcal{K}(\cdot)$ are the first-order modified hyperbolic Bessel functions \citep{abramowitz-stegun-1964} and $\xi=\sqrt{i2\pi /T}$.
Velocity profiles are shown in figure \ref{fig:gap-profile-oscill}. The Bessel layer was included in the code by reading in a map of the wall complex velocity at $t=0$. To advance in time the components within this map were multiplied by $e^{2\pi i {\hat t}/T}$ and the real components were extracted. As the boundary conditions are implemented in spectral space, it was necessary to Fourier transform the time-updated map of the velocity components at each time step, before passing the Fourier components as boundary conditions.

The difference between the values of drag reduction and power spent against the viscous forces computed by use of the two annular-gap models for $c=0,0.02D$, and $0.05D$ were within the uncertainty range estimated via numerical resolution checks based on variation of the mesh sizes, time step advancement, and size of the computational box (refer to RH13 for further details on the numerical resolution tests). For this reason and because of the higher computational cost caused by the Bessel profile due to the additional spectral transformations, the linear velocity profile model was used.  In order to choose the appropriate gap size for the simulations, the dimensional gap values were examined for typical experimental scenarios, presented in table 6 of RH13 for the steady disc flow case. The largest tested gap size of $c=0.05D$ was implemented as it corresponds to a value that would be achievable in the laboratory conditions detailed in this table.

\subsection{Flow decomposition}
\label{sec:disc-decomp}

The averaging operators used to decompose the flow are defined in the following. The space- and time-ensemble average is defined as
\begin{equation}
\label{ensemble-average}
\overline{f} (x,y,z,\tau)
=\frac{1}{N_x N_z N}
\sum_{n_x=0}^{N_x-1} 
\sum_{n_z=0}^{N_z-1}
\sum_{n_t=0}^{N-1}
f(\tilde{x}+2n_xD,y,\tilde{z}+n_zD,n_t T+\tau),
\end{equation}
where $2N_x$ and $N_z$ are the number of discs within the computational domain along $\tilde{x}$ and $\tilde{z}$, respectively, $\tau$ is the window time of the oscillation, and $N$ is the number of oscillation periods. 
The time average and the spatial average along the homogeneous directions are defined respectively as
\begin{equation}
\langle f \rangle (x,y,z)=\frac{1}{T}\int^{T}_0 \overline{f} (x,y,z,\tau) \mathrm{d}\tau, 
\quad
\hat{f}(y)=\frac{1}{L_xL_z}\int_0^{L_x}\int_0^{L_z} \langle f \rangle (x,y,z) \mathrm{d}z\mathrm{d}x\text{.}
\label{time-space-average}
\end{equation}
A global variable is defined as
\begin{equation*}
\left[f\right]_{g}=\int_0^{1}\hat{f}(y)\mathrm{d}y\text{.}
\end{equation*}
The size of all statistical samples is doubled by averaging over the two halves of the channel, taking into account the existing symmetries. The channel flow field is expressed by the sum
\begin{equation}
{\bf u}(x,y,z,t)={\bf u_m}(y)+{\bf u_d}(x,y,z,\tau)+{\bf u_t}(x,y,z,t),
\label{decomp}
\end{equation}
where ${\bf u_m}(y)=\{u_m,0,0\}=\hat{{\bf u}}$ is the mean flow, ${\bf u_d}(x,y,z,\tau)=\{u_d,v_d,w_d\}=\overline{{\bf u}} - {\bf u_m}$ is the disc flow, and ${\bf u_t}$ is the fluctuating turbulent component. 

\subsection{Performance quantities}
\label{sec:perfquants}
This section introduces the main quantities used to describe the oscillating-disc flow, i.e. the turbulent drag reduction, the power spent to activate the discs against the viscous resistance of the fluid, and the net power saved, which is their algebraic sum.

\subsubsection{Turbulent drag reduction}
The skin-friction coefficient $C_f$ is first defined as $C_f$$=$$2\tau_w^*/\left(\rho^*U_b^{* 2}\right)$, where $U_b^*$$=$$[u^*]_{g}/h^*$ is the bulk velocity. The latter is constant because the simulations are performed under conditions of constant mass flow rate. The drag reduction $\mathcal{R}$ is defined as the percentage change of the skin-friction coefficient with respect to the stationary wall value \citep{quadrio-ricco-2004}:  
\begin{equation}
\mathcal{R}(\%)=100 \frac{C_{f,s}-C_f}{C_{f,s}}\text{,}
\label{DR-1}
\end{equation}
where the subscript $s$ refers to the stationary wall case. Using $\tau_w^*=\mu^* u_m^{* \hspace{0.1mm} \prime}(0)$, where the prime denotes differentiation with respect to $y$, \eqref{DR-1} becomes $\mathcal{R}(\%)=100\cdot\left( 1-u_m^{\prime}(0)/u^{\prime}_{m,s}(0) \right)$.

\subsubsection{Power spent}
\label{sec:pspent}
As the oscillating disc flow is an active drag reduction technique, power is supplied to the system to move the discs against the viscous resistance of the fluid.  To calculate the power spent, term III of the instantaneous energy equation (1-108) in \citet{hinze-1975} is first considered. Its volume-average is the work done by the viscous stresses per unit time,
\begin{equation}
\mathcal{P}^*_{sp,t}=\frac{\nu^*}{L_x^*L_y^*L_z^*}\int^{L_x^*}_0\int_0^{L_y^*}\int^{L_z^*}_0\frac{\p}{\p x_i^*}\left[u_j^*\left(\frac{\p u_i^*}{\p x_j^*}+\frac{\p u_j^*}{\p x_i^*}\right)\right]\mathrm{d}\tilde{z}^*\mathrm{d}y^*\mathrm{d}\tilde{x}^*,
\label{hinze-psp}
\end{equation}
where $i,j$ are the indexes indicating the spatial coordinates $\tilde{x}$, $y$, $\tilde{z}$ and the corresponding velocity components (Einstein summation of repeated indexes is used). By substituting \eqref{decomp} into \eqref{hinze-psp} and by use of \eqref{ensemble-average} and \eqref{time-space-average}, one finds
\begin{equation}
\mathcal{P}^*_{sp,t}
=
\frac{\nu^*}{h^*}\left(\left. \widehat{u^*_d\frac{\p u^*_d}{\p y^*}}\right|_{y^*=0} + 
\left.\widehat{w^*_d\frac{\p w^*_d}{\p y^*}}\right|_{y^*=0}\right)\text{.}
\label{p-osc}
\end{equation}
The power spent \eqref{p-osc} is expressed as percentage of the power employed to drive the fluid in the streamwise direction, $\mathcal{P}^*_x$. By volume-, ensemble- and time-averaging the first term on the right-hand side of (1-108) in \cite{hinze-1975}, one obtains
\begin{equation}
\mathcal{P}^*_x=\frac{U^*_b \Pi^*}{\rho^*},
\label{p-x}
\end{equation}
By dividing \eqref{p-osc} by \eqref{p-x}, the percentage power employed to oscillate the discs with respect to the power spent to drive the fluid along the streamwise direction is obtained,
\begin{equation}
\mathcal{P}_{sp,t}(\%)= -\frac{100R_p}{R^2_\tau U_b}\left(\left.\widehat{u_d\frac{\p u_d}{\p y}}\hspace{-0.0mm}\right|_{y=0} + \left.\widehat{w_d\frac{\p w_d}{\p y}}\hspace{-0.0mm}\right|_{y=0}\right).
\label{pspent}
\end{equation}

\subsubsection{Net power saved}
\label{sec:pnet}
The net power saved, $\mathcal{P}_{net}$, the difference between the power saved due to the disc forcing (which coincides with $\mathcal{R}$ for constant mass flow rate conditions) and the power spent $\mathcal{P}_{sp,t}$, is defined as
\begin{equation}
\mathcal{P}_{net}(\%)=\mathcal{R}(\%)-\mathcal{P}_{sp,t}(\%).
\label{eq-pnet}
\end{equation}
 
\section{Laminar flow}
\label{sec:laminar}
For other active turbulent drag reduction techniques the analytical solutions for the corresponding laminar flows induced by wall motion have proven useful for accurately estimating important averaged turbulent quantities, such as the wall spanwise shear \citep{choi-xu-sung-2002}, the power spent for the wall forcing \citep{ricco-quadrio-2008}, and the thickness of the generalized Stokes layer generated by the wall waves \citep{skote-2011}. The laminar solution has also been employed to determine a scaling parameter which relates uniquely to drag reduction under specified wall forcing conditions \citep{quadrio-ricco-2004,cimarelli-etal-2013}. Through the laminar solution of the flow induced by a steadily rotating infinite disc, RH13 obtained an estimate of the time-averaged power spent to move the discs, which showed very good agreement with the power spent computed via DNS.

Inspired by previous works, the laminar flow above an infinite oscillating disc is therefore computed to calculate the power spent to activate the disc and to identify areas over the disc surface where the fluid performs work onto the discs, thus aiding the rotation. This is a form of the regenerative braking effect, studied by RH13 for steady disc rotation. These estimates are then compared with the turbulent quantities in \S\ref{sec:turbulent-pspent}.

\subsection{Laminar flow over an infinite oscillating disc}
\label{sec:lamsolver}
The laminar oscillating-disc flow was studied for the first time by \citet{rosenblat-1959} (refer to figure \ref{gap-geom} for the flow geometry). The velocity components are
\begin{align}
\{u_r^*,u_\theta^*\}=\frac{2r^* W^*}{D^*} \left\{F'\left(\eta,\breve{t}\right), G\left(\eta,\breve{t}\right)\right\},
\qquad 
u_y^*=-\frac{4 W^*}{D^*}\sqrt{\frac{\nu^* T^*}{\pi}}F\left(\eta,\breve{t}\right), 
\label{velocity-relations}
\end{align}
where the prime denotes differentiation with respect to $\eta=y^*\sqrt{\pi/(\nu^* T^*)}$, the scaled wall-normal coordinate, $\breve{t}= 2\pi t^*/T^*$ is the scaled time, and $u_r^*$, $u_\theta^*$ and $u_y^*$ are the radial, azimuthal, and axial velocity components, respectively. The following boundary conditions are satisfied
\begin{align*}
y^*=0:&\qquad u^*_r=0,\quad u^*_\theta=(2r^*W^*/D^*)\cos\breve{t}, \quad u^*_y=0, \quad p^*=0\text{.}\\
y^*\rightarrow\infty:&\qquad u^*_r=0, \quad u^*_\theta=0.
\end{align*}
Expressions \eqref{velocity-relations} are substituted into the cylindrical Navier-Stokes equations to obtain the equations of motion for $F'$ and $G$ under the boundary layer approximation,
\begin{align}
\begin{split}
\dot{F'}& = \frac{1}{2}F''' + \gamma(G^2+2FF''-F'^2)\text{,} \\
\dot{G}& = \frac{1}{2}G'' + 2\gamma(FG'-F'G),  
\end{split}
\label{eq:rosenblat-equations}
\end{align}    
with boundary conditions
\begin{equation}
\begin{array}{llll}
\eta=0: & F=F'=0, & G=\cos \breve{t}, \\ 
\eta\rightarrow\infty: & F'=G=0,
\label{fgh-bc-2}
\end{array}
\end{equation}
where the dot denotes differentiation with respect to $\breve{t}$ and $\gamma=T^*W^*/(\pi D^*)$. 
The latter parameter represents the ratio between the oscillation period $T^*$ and the period of rotation $\pi D^*/W^*$ which would occur if the disc rotated steadily with tip velocity $W^*$. The value $\gamma=\pi$ is relevant because it denotes the special case of maximum disc tip displacement equal to the circumference of the disc, i.e. each point at the disc tip covers a distance equal to $\pi D^*$ during a half period of oscillation.

The system \eqref{eq:rosenblat-equations}-\eqref{fgh-bc-2} was discretized using a first-order finite difference scheme for $\breve{t}$ and a second-order central finite difference scheme for $\eta$. The equations were first solved in time by starting from null initial profiles. The boundary condition for $G$ was altered as $G\left(0,\breve{t}\right)=1-e^{-\breve{t}}$ until $G$ was sufficiently close to unity. The system was then integrated with the boundary condition $G\left(0,\breve{t}\right)=\cos\breve{t}$. Figure \ref{G-profile} (left) shows the wall-normal profiles of $F'$ and $G$ at different oscillation phases.

\subsection{Laminar power spent}
\label{sec:laminar-pspent}

\begin{figure}
  \centering
  \includegraphics[width=\textwidth]{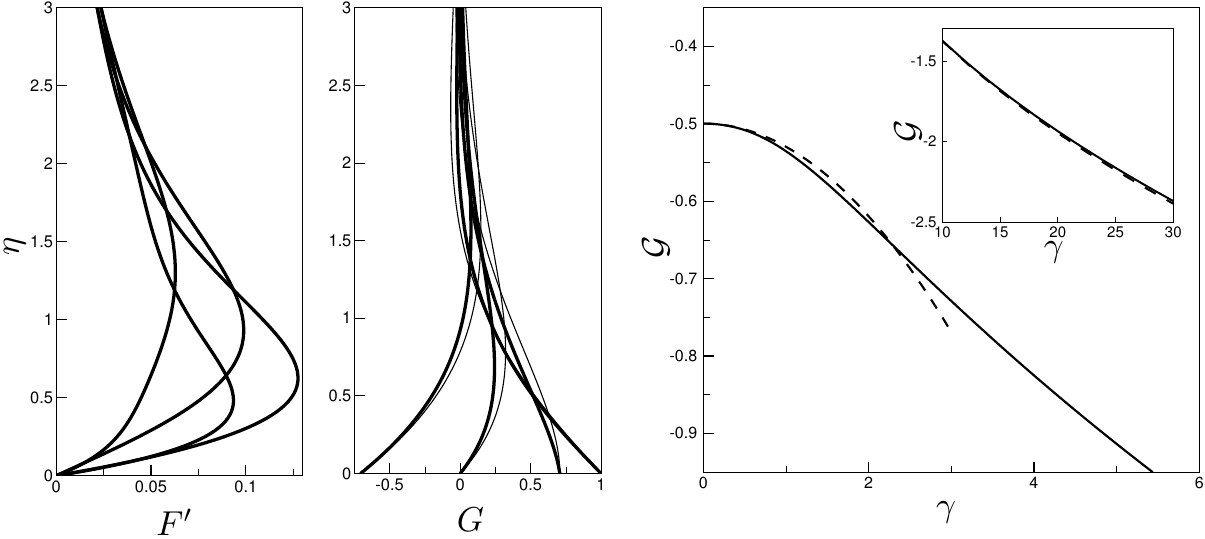}
  \caption{Left: Wall-normal profiles of $F'$ and $G$ at different oscillation phases for $\gamma=1$ (thick lines) and $\gamma=0$ (thin lines). The latter is given by \eqref{first-order-utheta} and coincides with the classical Stokes layer solution. Right: Numerically computed values of $\mathcal{G}(\gamma)$ (solid lines) and asymptotic solutions, \eqref{eq:GGp-asymptotic} for $\gamma \ll 1$ (dashed line in main plot), and \eqref{eq:G-large-gamma} for $\gamma \gg 1$ (dashed line in inset).}
  \label{G-profile}
\end{figure}
The laminar power spent $\mathcal{P}_{sp,l}^*$ is calculated using \eqref{hinze-psp}, where only ${\bf u_d}$ is retained in the laminar case as there is no mean streamwise flow above the disc and the turbulent fluctuations are null (${\bf u_m}={\bf u_t}=0$). Substituting $u_d=u_\theta\cos\theta$ and $w_d=u_\theta\sin\theta$ into \eqref{hinze-psp}, using \eqref{velocity-relations} and averaging over $\theta$, $r$, and time leads to
\begin{equation}
\mathcal{P}_{sp,l}^*=
\frac{\mathcal{G}(\gamma) W^{*\hspace{0.1mm}2}}{2}\sqrt{\frac{\pi\nu^{*\hspace{0.1mm}}}{T^{*\hspace{0.1mm}}}}\text{,}
\label{pspent-lam}
\end{equation}
where 
\begin{equation}
\mathcal{G}(\gamma) = \frac{1}{2\pi}\int^{2 \pi}_0 G\left(0,\breve{t}\right)G'\left(0,\breve{t}\right)  \mathrm{d}\breve{t}
\label{fancy-g}
\end{equation}
is shown in figure \ref{G-profile} (right). To express $\mathcal{P}_{sp,l}^*$ as percentage of the power spent to drive the fluid along the streamwise direction, \eqref{pspent-lam} is divided by \eqref{p-x} to obtain
\begin{equation}
\mathcal{P}_{sp,l}(\%)=\frac{50\mathcal{G}(\gamma)W^2 R_p^{3/2}}{U_b R_\tau^2}\sqrt{\frac{\pi}{T}}\text{.} 
\label{p-lam}
\end{equation}

\subsubsection{Asymptotic limit for $\gamma \ll 1$: the Stokes-layer regime}
\label{sec:gamma-small}

To obtain an analytical approximation to $\mathcal{G}$ for $\gamma \ll 1$, the expanded form of $G$ in powers of $\gamma$ can be used,
\begin{equation}
G_{\gamma \ll 1}(\eta,\breve{t},\gamma)=G_0(\eta,\breve{t})+\gamma^2G_2(\eta,\breve{t})+\mathcal{O}(\gamma^3)\text{,}
\label{eq:G-expanded}
\end{equation}
where $G_0$ and $G_2$ are given in equations (17) and (45) of \citet{rosenblat-1959}. Upon differentiation of \eqref{eq:G-expanded} with respect to $\eta$, the asymptotic form of $\mathcal{G}(\gamma)$ is
\begin{equation}
\label{eq:GGp-asymptotic}
\mathcal{G}_{\gamma \ll 1}(\gamma)
=
\frac{1}{2\pi}
\int_0^{2\pi}
G_0(0,\breve{t})
\left[G'_0(0,\breve{t})+\gamma^2G'_2(0,\breve{t})\right]
\mathrm{d}\breve{t}
=-\frac{1}{2}+\frac{\gamma^2}{160}\left(15\sqrt{2}-26\right) + \mathcal{O}(\gamma^3)\text{,}
\end{equation}
which is shown in figure \ref{G-profile} (right). The asymptotic solution predicts the numerical solution well for $\gamma<2$.

In the limit $\gamma \ll 1$, \citet{rosenblat-1959} obtained a first-order solution  
\begin{equation}
u_\theta^*=
\frac{2 r^*W^*}{D^*}e^{-\sqrt{\pi/(\nu^* T^*)}y^*}
\cos\left(\frac{2 \pi t^*}{T^*}-\sqrt{\frac{\pi}{\nu^* T^*}}y^*\right)\text{,}
\label{first-order-utheta}
\end{equation}
which is in the same form as the classical Stokes solution \citep{batchelor-1967}. Substituting \eqref{first-order-utheta} into \eqref{hinze-psp}, the first-order approximation is found, $\mathcal{P}_{sp,l}^*=0.25W^*\sqrt{\pi\nu^*/T^*}$, which is expressed as percentage of \eqref{p-x} to obtain
\begin{equation}
\mathcal{P}_{sp,l,\gamma \ll 1}=\frac{-25W^{2}R_p^{3/2}}{U_b R_\tau^2}\sqrt{\frac{\pi}{T}}\text{.}
\label{p-lam-gamma-small}
\end{equation}
This is also found directly from \eqref{p-lam} by setting $\mathcal{G}(0) =-0.5$.

\subsubsection{Asymptotic limit for $\gamma \gg 1$: the quasi-steady regime}
\label{sec:gamma-large}

As suggested by \citet{benney-1964}, in the limit $\gamma \gg 1$ it is more appropriate to rescale the wall-normal coordinate by the Ekman layer thickness $\delta_e^*=\sqrt{\nu^* D^*/(2W^*)}$. The rescaled equations (2.19) and (2.20) of \citet{benney-1964} were then solved using the same numerical method described in \S\ref{sec:lamsolver}. The von K\'arm\'an equations describing the flow over a steadily rotating disc are recovered in the limit $\gamma \rightarrow \infty$. The asymptotic limit of $\mathcal{G}$ for $\gamma \gg 1$ is found by first rescaling $G'(0,\breve t)$ in \eqref{fancy-g} through $\delta_e^*$ and by noting that the time modulation of the disc motion enters the problem only parametrically, 
\begin{equation}
\label{g-prime-asy}
G'_{\gamma \gg 1}(0,\breve t) = \sqrt{2 \gamma} G_s \cos{\breve t},
\end{equation}
where $G_s=-0.61592$ \citep{rogers-lance-1960}. By substituting \eqref{g-prime-asy} into \eqref{fancy-g} and by use of \eqref{fgh-bc-2}, one finds
\begin{equation}
\label{eq:G-large-gamma}
\mathcal{G}_{\gamma \gg 1}(\gamma) = G_s \sqrt{\frac{\gamma}{2}} \text{.}
\end{equation}
As shown in figure \ref{G-profile} (right, inset), the asymptotic expression \eqref{eq:G-large-gamma} matches the numerical values well. By substituting \eqref{eq:G-large-gamma} into \eqref{p-lam}, the asymptotic form of the power spent is obtained
\begin{equation}
\mathcal{P}_{sp,l,\gamma \gg 1}=\frac{50 G_s W^{3/2} R_p^{3/2}}{\sqrt{2 D} U_b R_\tau^2}\text{.}
\label{p-lam-gamma-large}
\end{equation}
By coincidence, the power spent when $\gamma=0$, i.e. \eqref{p-lam-gamma-small}, is half of the oscillating-wall case at the same $W^*$ and $T^*$ \citep{ricco-quadrio-2008}, and the power spent when $\gamma \gg 1$, i.e. \eqref{p-lam-gamma-large}, is half of the steady-rotation case at the same $W^*$ and $D^*$ (RH13). The oscillating-disc power spent is expected to be smaller than in these two cases, but for different reasons. The oscillating-wall case requires more power because the motion involves the entire wall surface, while the steady-rotation case consumes more power because the motion is uniform in time.

\subsection{Laminar regenerative braking effect}
\label{app-lampower}
The laminar phase- and time-averaged power spent $\mathcal{W}_l$ to oscillate the discs beneath a uniform streamwise flow is computed by following RH13. As the purpose of this analysis is to obtain a simple estimate of the turbulent case, the streamwise shear flow is superimposed on the Rosenblat flow without considering their nonlinear interaction. A rigorous study of this flow would be the extension of the work by \citet{wang-1989} with oscillatory wall boundary conditions. Starting from \eqref{hinze-psp}, using \eqref{decomp}, and setting ${\bf u_t}=0$, one finds
\begin{equation}
\mathcal{W}_l(x,0,z,\breve{t})
=
\frac{1}{R_p}\left[ u_d(x,0,z,\breve{t})\left( u_m'(0) + \left.\frac{\p u_d}{\p y}\right|_{y=0} \right) + w_d(x,0,z,\breve{t}) \left.\frac{\p w_d}{\p y}\right|_{y=0}\right]\text{.}
\label{plam-space-1}
\end{equation}
Using \eqref{velocity-relations}, \eqref{plam-space-1} becomes
\begin{equation*}
\mathcal{W}_l(r,\breve{t})
=
\frac{2rWG(0,\breve{t},\gamma)}{D R_p}
\left( 
u_m'(0)\cos\theta + \frac{2Wr}{D}\sqrt{\frac{\pi R_p}{T}}G'(0,\breve{t},\gamma) 
\right).
\label{plam-space-2}
\end{equation*}
By rearranging to obtain an inequality in $r$, the region where the streamwise flow exerts work on the disc (regenerative braking effect) is found,
\begin{equation}
r<-\frac{u'_m(0) D\cos\theta}{2WG'(0,\breve{t},\gamma)}\sqrt{\frac{T}{\pi R_p}}\text{.}
\label{plam-space-3}
\end{equation}
In \S\ref{sec:turbulent-pspent}, the region of regenerative braking effect is computed for the turbulent case and compared with the laminar prediction \eqref{plam-space-3}.
 
\section{Turbulent flow}
\label{sec:turbulent}

The turbulent flow results are presented in this section. Sections \S\ref{sec:turbulent-time}, \S\ref{sec:turbulent-dependence}, \S\ref{sec:turbulent-FIK}, \S\ref{sec:turbulent-scaling-s} focus on the drag reduction, section \S\ref{sec:turbulent-discflow} presents disc flow visualization and statistics, and section \S\ref{sec:turbulent-pspent} describes the power spent to move the discs and the comparison with the laminar prediction, studied in \S\ref{sec:laminar-pspent}.

\subsection{Time evolution}
\label{sec:turbulent-time}
\begin{figure} 
\centering                                                                                                                                                                                                                    
\includegraphics[width=\textwidth]{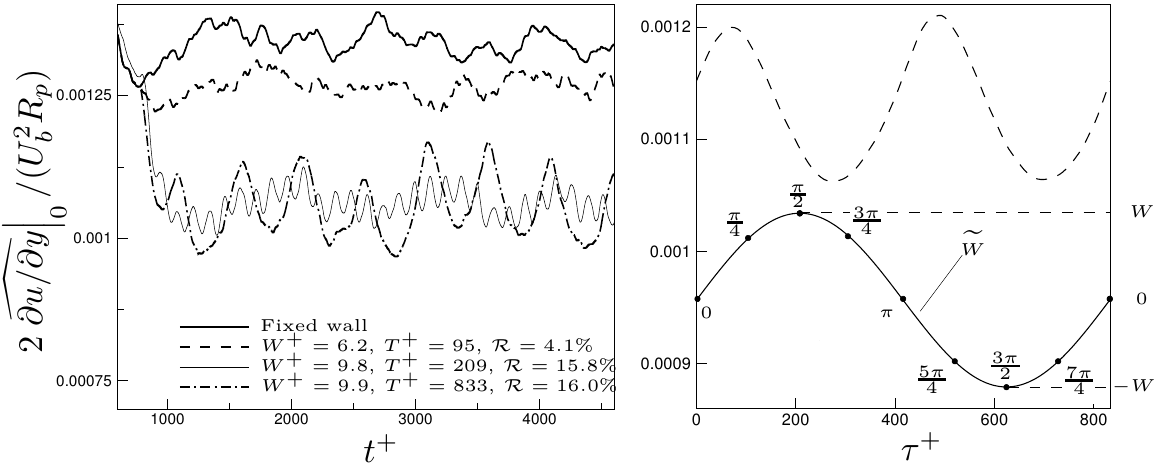} 
\caption{Left: Space-averaged streamwise wall-shear stress vs. time for cases at $D=3.38$. The disc forcing is initiated at $t^+=770$. Only a fraction of the total integration time is shown. The space-averaging operator here does not include time averaging. Right: Ensemble- and space-averaged streamwise wall-shear stress vs. $\tau^+$ for $D^+=554$, $W^+=9.9$, $T^+=833$ (dashed line). The disc velocity is shown by the solid line. The phase $\phi$ is given in the figure.}
\label{D640-hist}
\end{figure}

The temporal evolution of the space-averaged wall-shear stress is displayed in figure~\ref{D640-hist} (left). The transient time occurring between the start-up of the disc forcing and the fully established disc-altered regime increases with $\mathcal{R}$. This agrees with the oscillating wall and RH13, but the duration of the transient for the discs is shorter than for the oscillating wall case. The time modulation of the wall-shear stress is notable for the high $\mathcal{R}$ cases, with the amplitude of the signal increasing with $T$. The significant time modulation and the shorter transient compared with the oscillating wall technique could be due to the discs forcing the wall turbulence in the streamwise direction. 
The streamwise wall-shear stress is therefore affected directly whereas in the oscillating-wall case the streamwise shear flow is modified indirectly as the motion is along the spanwise direction only. 

The space- and phase-averaged wall-shear stress modulation, shown by the dashed line in figure \ref{D640-hist} (right), has a period equal to half of the wall velocity. This is expected because of symmetry of the unsteady forcing with respect to the streamwise direction. The wall-shear stress reaches its minimum value approximately $T/8$ after the disc velocity is maximum, i.e. at $\phi=5\pi/8, 13\pi/8$.  The wall-shear stress peaks approximately $T/8$ after the disc velocity is null, i.e. at $\phi=\pi/8, 9\pi/8$.

\subsection{Dependence of drag reduction on $D$, $W$, $T$}
\label{sec:turbulent-dependence}
\begin{figure}
   \centering
   \includegraphics[width=\textwidth]{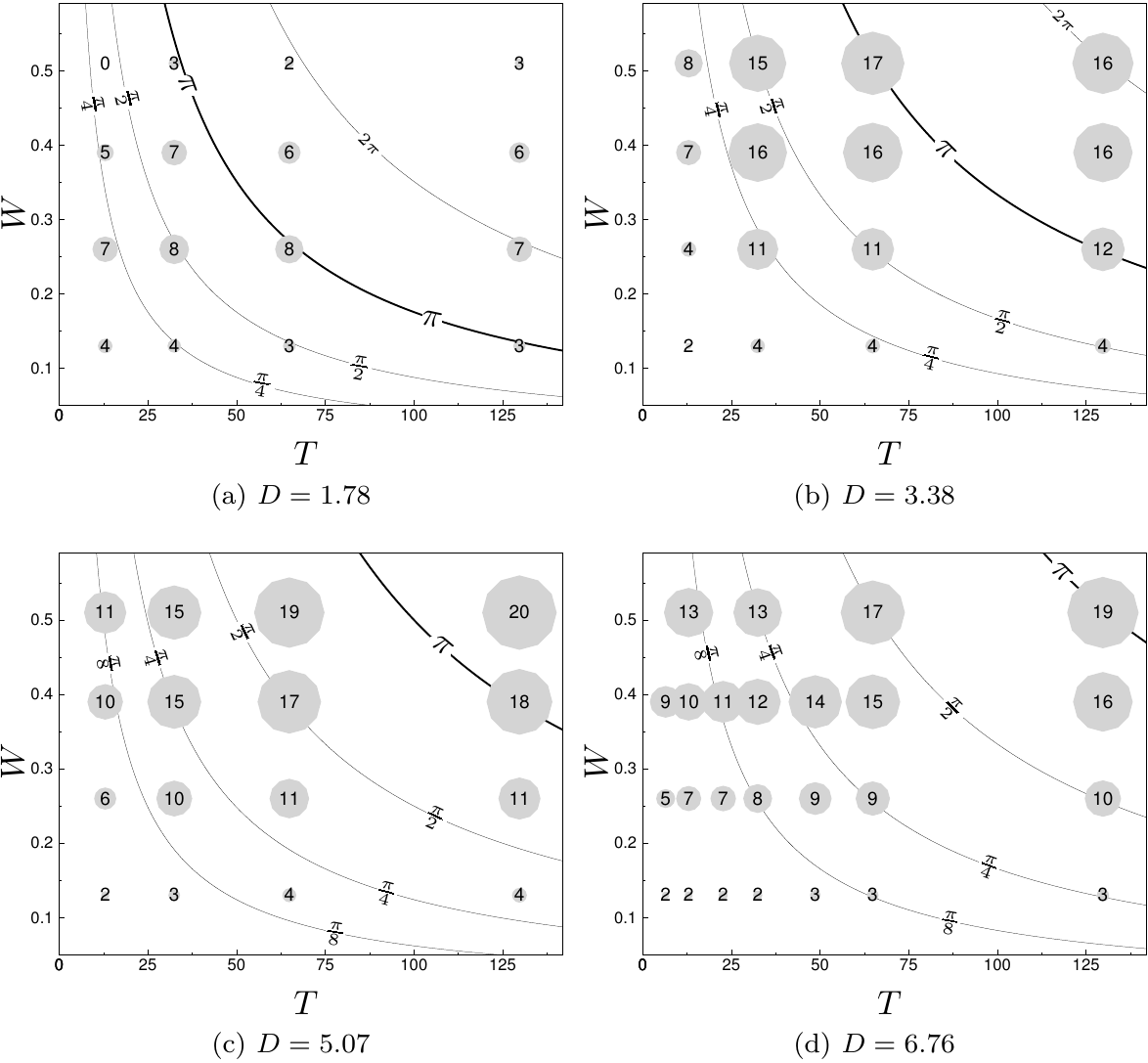}
   \caption{Plots of $\mathcal{R}(T,W)(\%)$ for different $D$. The circle size is proportional to the drag reduction value. The hyperbolae are constant-$\gamma$ lines.}
   \label{D-hr-outer}
\end{figure}

\begin{figure}
   \centering
   \includegraphics[width=\textwidth]{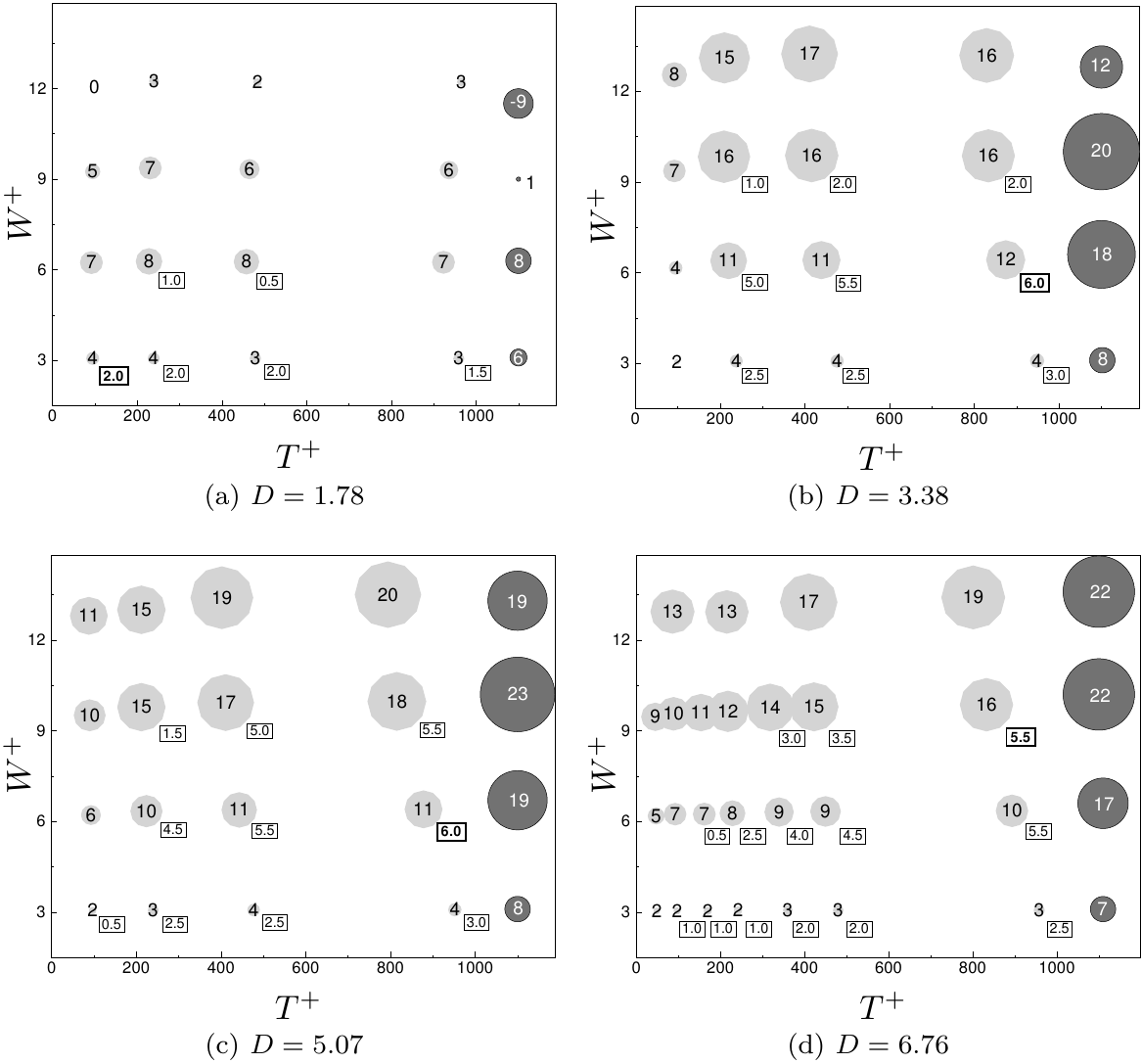}
   \caption{Plots of $\mathcal{R}(T^+,W^+)(\%)$. Scaling is performed using $u_\tau^*$ from the native case. The dark grey circles indicate RH13's data and the boxed values denote positive $P_{net}$ values.}
   \label{D-hr}
\end{figure}

Figure \ref{D-hr-outer} depicts maps of $\mathcal{R}(T,W)(\%)$ for disc sizes $D=1.78$, $3.38$, $5.07$, and $6.76$. The $\gamma$ values are shown as hyperbolae in these planes. For cases with $\gamma>\pi$, the maximum displacement is larger than the disc circumference.
Figure \ref{D-hr} shows the same drag-reduction data, scaled in viscous units. The boxed values represent the net power saved $\mathcal{P}_{net}(\%)$ defined in \eqref{eq-pnet}. Only positive $\mathcal{P}_{net}$ values are shown and the bold boxes highlight the maximum $\mathcal{P}_{net}$ values.

For $D=1.78$ and 3.38 and fixed $W$, drag reduction increases up to an optimum $T$ beyond which it decays. This optimum $T$ depends on $D$, and increases with the disc diameter. For $D=1.78,3.38$ the optimal periods are in the ranges $T^+=200-400$ and $T^+=400-800$, respectively. For $D=5.07$ and $6.76$ the optimal period is not computed and therefore $\mathcal{R}$ increases monotonically with $T$ for fixed $W$ and $D$. Cases with larger $T$ are not investigated due to the increased simulation time required for the averaging procedure. 

For $D=1.78$ and fixed $T$, drag reduction increases up to an optimum wall velocity, $W \approx 0.26$ ($W^+ \approx 6$), above which drag reduction decreases. This behaviour also occurs in the steady-disc case studied by RH13. The optimal $W$ are not found for larger $D$ as the drag reduction increases monotonically with $W$ for fixed $D$ and $T$.

For $T \gg 1$, the wall forcing is quasi-steady and it is therefore worth comparing the $\mathcal{R}$ value with the ones obtained by steady disc rotation, computed by RH13. RH13's values are however not expected to be recovered in this limit. A primary reason for this is that the power spent in the oscillating-disc case is smaller than in the steady rotation case, as verified in \S\ref{sec:turbulent-pspent} (in \S\ref{sec:gamma-large}, it is predicted to be half of the steady case by use of the laminar solution when the oscillation period is large). RH13's values are displayed in figure \ref{D-hr} by the dark grey circles on the right-hand side of each map. In most of the cases where the optimal $T^+$ is detected, i.e. for $W^+>3$, $D=1.78$, and for $W^+>9$, $D=3.38$ and 5.07, our $\mathcal{R}$ values may reach larger values than RH13's for the same $W^+$. For $D=6.76$, all our computed $\mathcal{R}$ are lower than RH13's. 

Figure \ref{D-hr} also shows that a positive $\mathcal{P}_{net}$ occurs only for $W^+\leq9$. This confirms the finding by RH13 for steady rotation and is expected because the power spent grows rapidly as $W$ grows, as also suggested by the laminar result in \eqref{p-lam}. The largest positive $\mathcal{P}_{net}$ in the parameter range is $6\pm1\%$, and is obtained for $D^+=855$, $W^+=6.4$, $T^+=880$, and $D^+=568$, $W^+=6.4$, $T^+=874$.


\subsection{The Fukagata-Iwamoto-Kasagi identity}
\label{sec:turbulent-FIK}
The Fukagata-Iwamoto-Kasagi (FIK) identity relates the skin-friction coefficient of a wall-bounded flow to the Reynolds stresses \citep{fukagata-iwamoto-kasagi-2002}. It is extended here to take into account the oscillating-disc flow effects (the reader should refer to Appendix A of RH13 for a slightly more detailed derivation for the steady disc flow case). By non-dimensionalizing the streamwise momentum equation into outer units, decomposing the velocity field as discussed in \S\ref{sec:disc-decomp} and averaging in time, along the homogeneous $x$ and $z$ directions, and over both halves of the channel, the following is obtained
\begin{equation*}
\Pi Re_p= \left( u_m^\prime - \widehat{u_dv_d} - \widehat{u_t v_t} \right)^\prime\text{,}
\label{FIK-anal-1}
\end{equation*}
where the prime indicates differentiation with respect to $y$. By following the same procedure outlined in \citet{fukagata-iwamoto-kasagi-2002} and noting that the Reynolds stresses term $\widehat{u_t v_t}$ in equation (1) in \citet{fukagata-iwamoto-kasagi-2002} is replaced with the sum $\widehat{u_t v_t} + \widehat{u_d v_d}$, the relationship between $C_f$ and the Reynolds stresses for the disc flow case can be written as
\begin{equation}
C_f=\frac{6}{U_bRe_p}-\frac{6}{U_b^2}\left[\left(1-y\right)\left( \widehat{u_t v_t} + \widehat{u_d v_d} \right) \right]_g\text{,}
\label{FIK-outer}
\end{equation}
which is in the same form of the steady case by RH13.
The drag reduction computed through the Reynolds stresses via \eqref{FIK-outer} is $\mathcal{R}=16.9\%$ for $D=3.38$, $W^+=13.2$ and $T^+=411$, which agrees with $\mathcal{R}=17.1\%$, calculated via the wall shear-stress. Using \eqref{FIK-outer}, it is also possible to separate the total drag reduction into the change of the turbulent Reynolds stresses $\widehat{u_tv_t}-\langle\widehat{u_{t,s}v_{t,s}}\rangle$ and the contribution of the time averaged disc Reynolds stresses $\widehat{u_dv_d}$, i.e. $\mathcal{R}(\%)=\mathcal{R}_t(\%)+\mathcal{R}_d(\%)$ where
\begin{align}
\label{rt}
\mathcal{R}_t(\%)&=100\frac{R_p\left[\left(1-y\right)\left(\widehat{u_t v_t}-\langle\widehat{u_{t,s}v_{t,s}}\rangle\right) \right]_g}{U_b-R_p\left[\left( 1-y\right)\langle\widehat{u_{t,s}v_{t,s}}\rangle\right]_g}\text{,} \\
\quad
\label{rd}
\mathcal{R}_d(\%)&=100\frac{R_p\left[\left(1-y\right)\widehat{u_dv_d}\right]_g}{U_b-R_p\left[\left(1-y\right)\langle\widehat{u_{t,s}v_{t,s}}\rangle\right]_g}\text{.}
\end{align}
The subscript $s$ again refers to the stationary wall case. This decomposition is used in section \S\ref{sec:turbulent-scaling-s} to study the drag reduction physics.

\subsection{Disc flow visualisations and statistics}
\label{sec:turbulent-discflow}
\begin{figure}
\centering
\includegraphics[width=\textwidth]{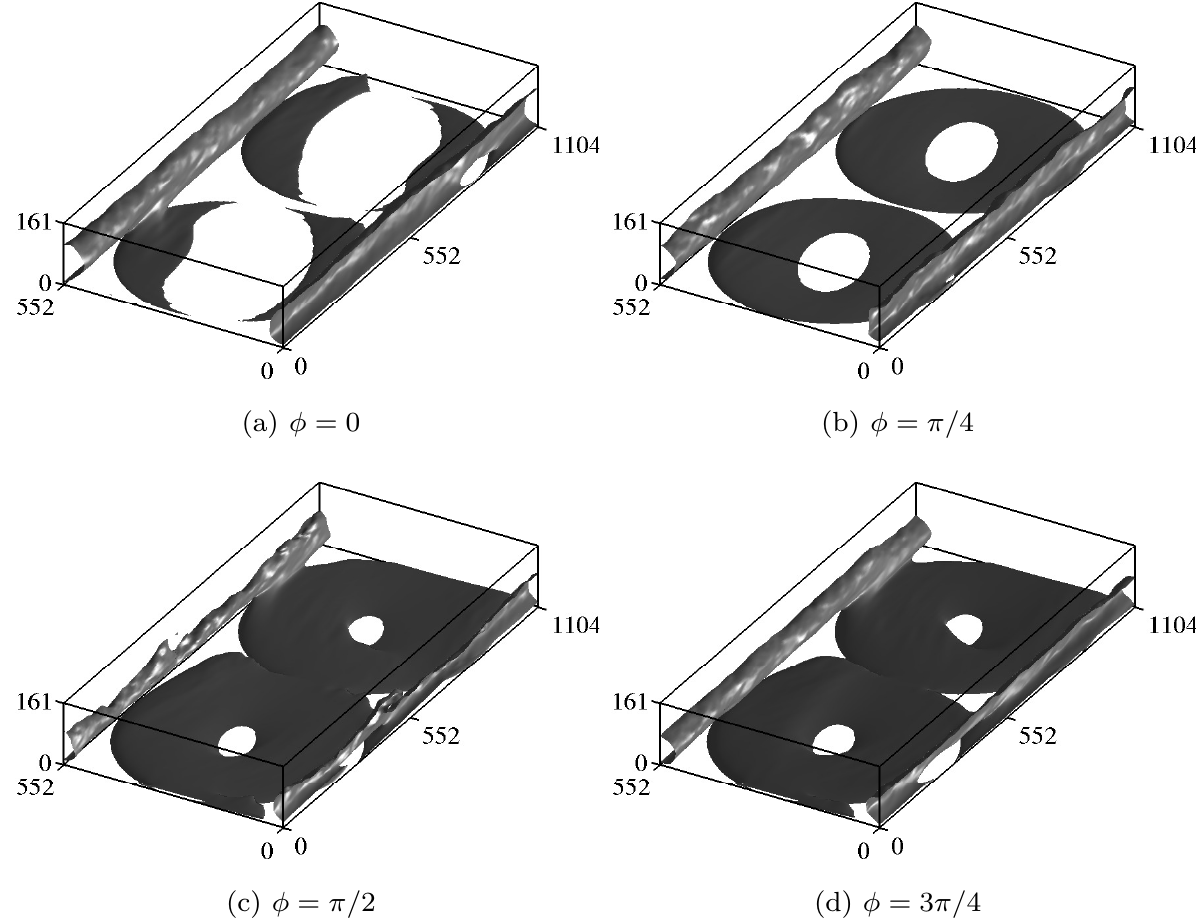}
\vspace{0.25cm}
\caption{Disc-flow visualisations of $q^+(x,y,z)=\sqrt{u_d^{+\hspace{0.1mm}2}+w_d^{+\hspace{0.1mm}2}}=2.1$ at phases $\phi=0,\pi/4,\pi/2,3\pi/4$. The disc tip velocity at each phase is shown in figure \ref{D640-hist} (right). In this figure and in figures \ref{ubands}, \ref{cf-contours}, \ref{discflow}, and \ref{pspent-space}, $D^+=552$, $W^+=13.2$, $T^+=411$.}
\label{discflow-vis}
\end{figure}

The disc flow for $D^+=552$, $W^+=13.2$ and $T^+=411$ ($\mathcal{R}=17\%$) is visualized at different phases in figure \ref{discflow-vis}. Isosurfaces of $q^+=\sqrt{u_d^{+2}+w_d^{+2}}=2.1$ are displayed. Similarly to the steady case by RH13, streamwise-elongated tubular structures appear between discs, which extend vertically up to almost one quarter of the channel height. They occur where there is high tangential shear, i.e. where the disc tips are next to each other and rotate in opposite directions, but also over sections of stationary wall. They persist almost undisturbed across the entire period of oscillation, their intensity and shape being only weakly modulated in time. The thin circular patterns on top of the discs instead show a strong modulation in time. This is expected as the patterns are directly related to the disc wall motion. Although at $\phi=0$ the disc velocity is null, the circular patterns are still observed as the rotational motion has diffused upward from the wall by viscous effects.  
Instantaneous isosurfaces of low-speed streaks in the proximity of the wall (not shown) reveal that the intensity of these structures is weakened significantly, similarly to the steady disc-flow case. 

\begin{figure}
\centering
\includegraphics[width=\textwidth]{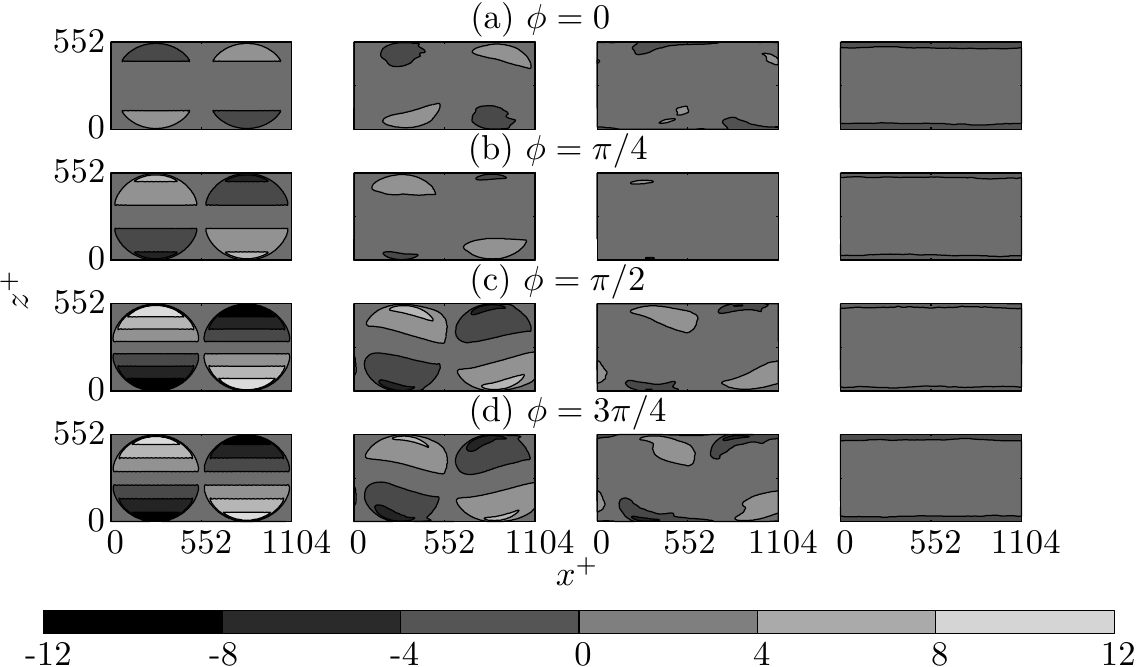}
\vspace{0.25cm}
\caption{Contour plot of $u_d^+(x,y,z)$ as a function of phase in the $x-z$ plane at $y^+=0$, $y^+=4$, $y^+=8$ and $y^+=27$ (from left to right).}
\label{ubands}
\end{figure}

Contour plots of $u_d$ in $x-z$ planes are shown in figure \ref{ubands}. The first column on the left shows the contour at the wall. At $y^+=4$ and $y^+=8$, the disc outlines can still be observed, the clarity decreasing with the  increased distance from the wall. At these heights the contour lines are no longer straight, but show a wavy modulation. The circular patters created by the disc motion are displaced in the streamwise direction by the mean flow. The magnitude of the shift increases with distance from the wall and at $y^+=8$ it is about $100\nu^*/u_\tau^*$. At $y^+=27$ the disc outlines are no longer visible and the structures occurring between discs in figure \ref{discflow-vis} here appear as streamwise-parallel bands of $u_d$ which do not modulate in time and are slower than the mean flow. They also appear at higher wall-normal locations up to the channel half-plane, with their width increasing with height.

\begin{figure}
\includegraphics[width=1.0\textwidth]{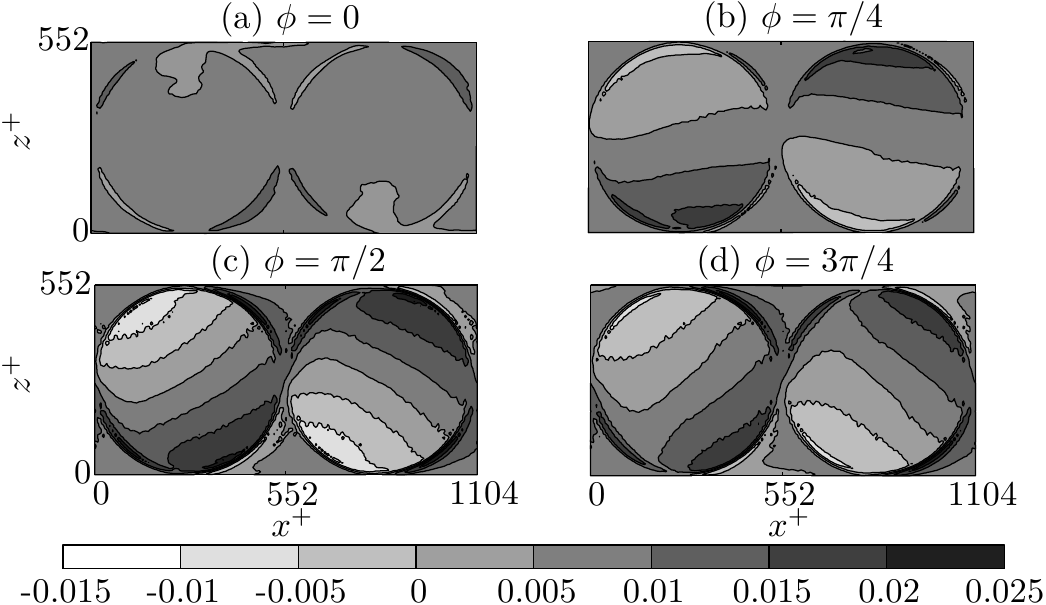}%
\caption{Contour plot of phase-averaged streamwise wall friction, $2\left.\langle\p u^+/\p y^+\right|_0\rangle/U_b^{+\hspace{0.1mm}2}$. The skin-friction coefficient is $C_f=6.79\cdot10^{-3}$.}
\label{cf-contours}
\end{figure}

The contour plots in figure \ref{cf-contours} show the ensemble- and time-averaged wall-shear stress. At phases $\phi=0$ and $\pi$, when the angular velocity of the discs is zero, the wall-shear stress is almost uniform over the disc surface. During the other phases of the cycle, the lines of constant stress are inclined with respect to the streamwise direction and the maximum values are found near the disc tip. The lines show a maximum inclination of about $45^\circ$ at phases $\phi=3\pi/4,7\pi/4$, when the deceleration of the discs is maximum.

\begin{figure}%
\centering
\includegraphics[width=\textwidth]{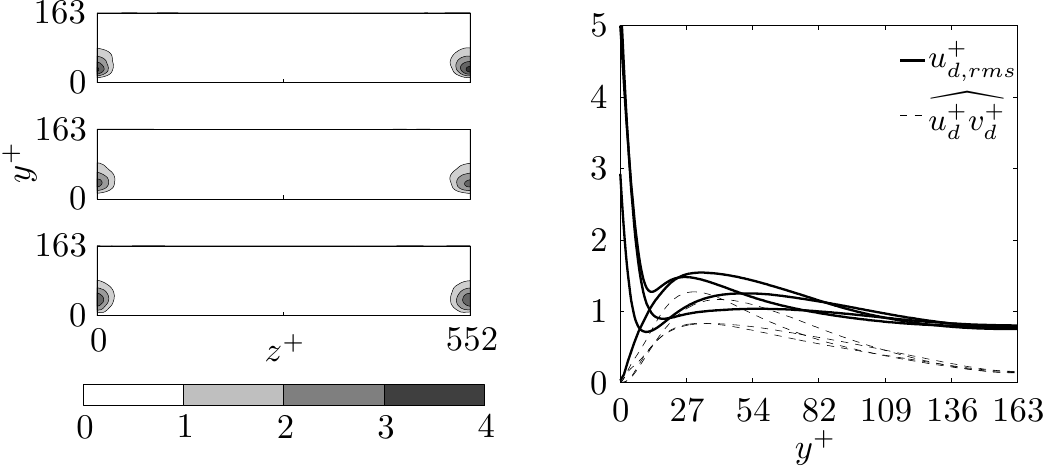}
\vspace{0.25cm}
\caption{Left: Isosurfaces of $\langle u_d^+v_d^+\rangle$ observed from the $y-z$ plane at $x^+=0$, $x^+=160$, $x^+=320$ (from left to right). The plot shows only $\langle u_d^+v_d^+\rangle$ for $u_d,v_d<0$ as within the contour range the contributions from other combinations of $u_d$ and $v_d$ are negligible.  Right: Wall-normal profiles of the $u_{d,rms}^+$ (solid lines) and $\widehat{u_d^+ v_d^+}$ (dashed lines). Profiles are shown for phases from the first half of the disc oscillation.}
\label{discflow}
\end{figure}

Figure \ref{discflow} (left) shows contours of the time-averaged $\langle u_dv_d\rangle$ observed from the $y-z$ plane at different streamwise locations. These contours overlap with the elongated structures in figures \ref{discflow-vis} and \ref{ubands}, which are therefore recognized as primarily responsible for these additional Reynolds stresses. It is clear that the structures are only slowly varying along the streamwise direction. The flow over the disc surface does not contribute to $\langle u_dv_d\rangle$ because, although $u_d$ is significant, $v_d$ is negligible. Only the contribution to $\langle u_dv_d\rangle$ from both negative $u_d$ and $v_d$ is included in figure \ref{discflow} (left) as $u_d$ and $v_d$ with other combinations of signs only negligibly add to the total stress. The structures are therefore jets oriented toward the wall and backward with respect to the mean flow.

Figure~\ref{discflow} (right) shows the time modulation of the root-mean-square (r.m.s.) of the disc streamwise velocity component, defined as $u_{d, rms}(y,\tau)=\sqrt{\widehat{u_d^2}}$, and of the Reynolds stresses $\widehat{u_d^+ v_d^+}$ (where here the spatial average $\widehat{\cdot}$ does not include the time average as in \eqref{time-space-average}). Four profiles are shown for each quantity, for phases from the first half period of the oscillation. Data from the second half are not shown as the profiles coincide at opposite oscillation phases. The disc flow penetrates into the channel up to $y^+\approx15$. When the disc tip velocity is close to its maximum, the profiles of $u_{d,rms}$ and $w_{d,rms}$ (the latter not shown) decay from their wall value and follow each other closely up to $y^+\approx10$. At higher locations, the magnitude of $u_{d,rms}^+$  is larger than that of the wall-normal and spanwise velocity profiles.  
In the bulk of the channel, for $y^+>50$, the profiles modulate only slightly in time. This therefore further confirms that the intense temporal modulation of the disc flow is confined in the viscous sublayer and buffer region.  $u_{d,rms}^+$ decays to $\approx0.7$ as the channel centreline is approached. As expected, the Reynolds stresses $\widehat{u_d^+ v_d^+}$  show a slow time modulation and are always positive, proving that the streamwise-elongated structures favourably contribute to the drag reduction through $\mathcal{R}_d$ in 
\eqref{rd}. Neither $u_{d,rms}^+$ nor $\widehat{u_d^+ v_d^+}$ modulate in time for $y^+>120$. 

\subsection{Power spent}
\label{sec:turbulent-pspent}

\subsubsection{Comparison with laminar power spent}
\label{sec:comparison}

\begin{figure} 
  \centering
  \includegraphics[width=\textwidth]{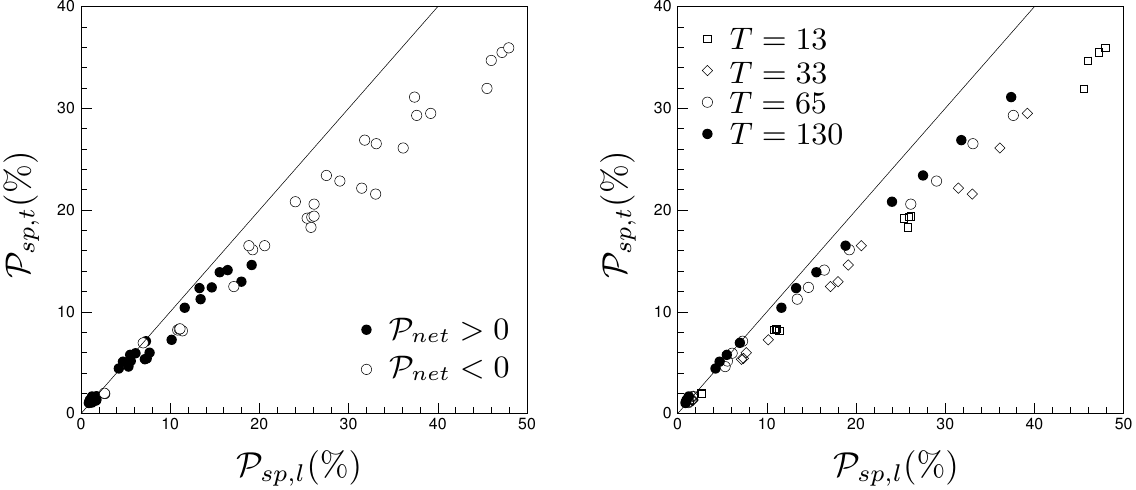}
  \caption{Left: $\mathcal{P}_{sp,t}(\%)$, computed through DNS via \eqref{pspent}, vs. $\mathcal{P}_{sp,l}(\%)$, computed through \eqref{p-lam}, the power spent by an infinite disc oscillating beneath a still fluid. Data are coloured according to $\mathcal{P}_{net}$. Right: $\mathcal{P}_{sp,t}(\%)$ vs. $\mathcal{P}_{sp,l}(\%)$, with symbols grouped according to $T$. 
}
\label{pspent-grouped}
\end{figure}

Figure \ref{pspent-grouped} (left) shows the comparison between the power spent $\mathcal{P}_{sp,t}$ to impose the disc motion, computed via \eqref{pspent} with DNS data, and the laminar power spent, calculated via \eqref{p-lam}. The values match satisfactorily for low $\mathcal{P}_{sp,t}$, and the disagreement grows for larger $\mathcal{P}_{sp,t}$. This is due to the larger values of $W$, which intensify the nonlinear interactions between the disc flow and the streamwise turbulent mean flow, and promote the interference between neighbouring discs. As the laminar calculations are performed by not accounting for the disc interference through the assumption of infinite disc size and by neglecting the streamwise mean flow, the agreement is expected to worsen for large $W$. Figure \ref{pspent-grouped} (left) also shows that the power spent for cases with positive $\mathcal{P}_{net}$ is predicted more accurately by the laminar solution than for cases with negative $\mathcal{P}_{net}$, a result also found by RH13.

Figure~\ref{pspent-grouped} (right) presents the same data of the right plot, with the symbols coloured according to $T$. The agreement is best for the largest oscillation periods, $T=130$, and it worsens as $T$ decreases. The trend for $T=130$ closely resembles the one of the steadily rotating discs by RH13, which is consistent with the wall forcing becoming quasi-steady at large periods. For $T=130$, the highest value of $\mathcal{P}_{sp,t}=37\%$, occurring for $D=1.78$, $W=0.51$, differs from $\mathcal{P}_{sp,l}$ by 17\%, while a disagreement of 15\% is found by RH13 for the same $\mathcal{P}_{sp,l}$ value. 

\subsubsection{Turbulent regenerative braking effect}
\label{sec:turbulent-regenerative}

\begin{figure}
\centering
\includegraphics[width=\textwidth]{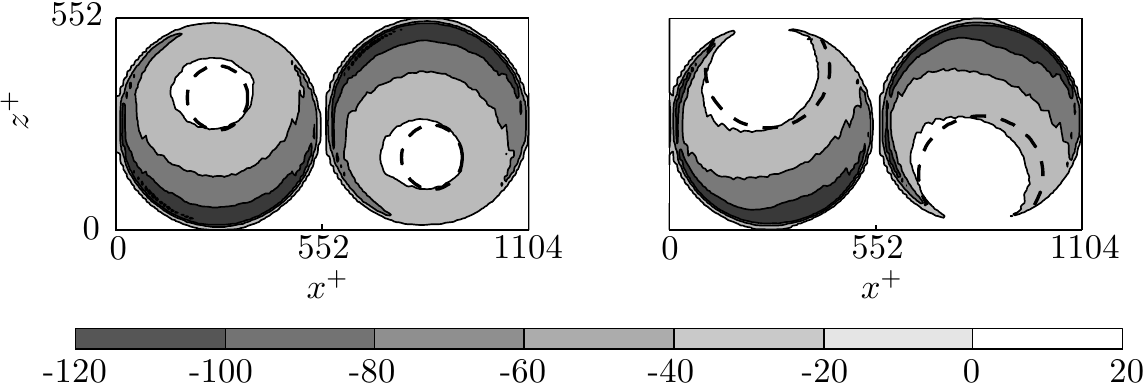}
\caption{Spatial variation of $\mathcal{W}_t$, computed via \eqref{psp-tau}, for $\phi=\pi/4$ (left) and $\phi=3\pi/4$ (right).  The white areas over the disc surfaces for which $\mathcal{W}_t>0$ denote locations where the fluid is performing work onto the disc. The areas of regenerative braking predicted by the laminar solution, i.e. where $\mathcal{W}_l>0$ and \eqref{plam-space-3} applies, are enclosed by the dashed lines.}
\label{pspent-space}
\end{figure}

For the majority of oscillation cycle, power is spent by the discs to overcome the frictional resistance of the fluid. However, for part of the oscillation and over portion of the disc surface, work is performed by the fluid on the disc. This is a form of regenerative breaking effect and it also occurs in time for the case of uniform spanwise wall oscillations and in space for the steady rotating disc case (RH13). Contour plots of the localized power spent $\mathcal{W}_t$, defined as
\begin{equation}
\mathcal{W}_t(x,z,\tau)(\%)=\frac{100R_p}{R_\tau^2 U_b}
\left(\hspace{0.2mm} 
\left.u_d\frac{\p u_d}{\p y}\right|_{y=0} 
+  
\left.w_d\frac{\p w_d}{\p y}\right|_{y=0}
\hspace{0.2mm}\right)
\text{,}
\label{psp-tau}
\end{equation}
are shown in figure \ref{pspent-space} for $\phi=\pi/4,3\pi/4$. The white regions over the disc surface correspond to the regenerative braking effect, where $\mathcal{W}_t \geq 0$, i.e. the fluid performs work on the discs. The dashed lines represent the regions of $\mathcal{W}_l(r,\tau)>0,$ predicted through the laminar solution by \eqref{plam-space-3}. Although the regenerative braking areas computed via DNS are slightly shifted upstream when compared with those predicted through the laminar solution, the overall agreement is very good and better than in RH13's case.  

\subsection{A discussion on drag reduction physics and scaling}
\label{sec:turbulent-scaling-s}

\begin{figure}
\centering
\includegraphics[width=0.75\textwidth,trim=0 6cm 0 0,clip=true]{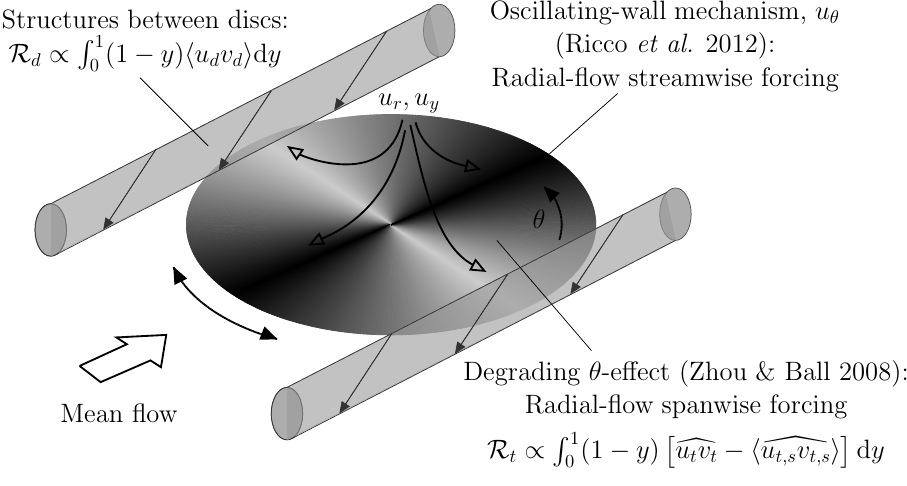}
\caption{Schematic of the two mechanisms responsible for drag reduction induced by oscillating discs. One mechanism is linked to the attenuation of the turbulent Reynolds stresses and is quantified by $\mathcal{R}_t$ in \eqref{rt}. The degrading effect of the oscillation angle $\theta$ \citep{zhou-ball-2006} is represented by the shading. The second mechanism is due to the structures between discs and is quantified by $\mathcal{R}_d$ in \eqref{rd}.  The radial streaming induced by the Rosenblat pump is denoted by the open arrows.} 
\label{schematic}                  
\end{figure}

The results in the preceding sections prove that the oscillating discs effectively modify the flow in two distinct ways, which are discussed in the following and illustrated in figure \ref{schematic}.

\begin{itemize}
\item {\em Role of disc boundary layer} 

The circular pattern which forms over a disc as a direct consequence of the disc rotation (shown in figure \ref{discflow-vis}) is a thin region of high-shear flow. The laminar analysis suggests that this oscillatory boundary layer resembles the oscillating-wall Stokes layer (of thickness $\delta^*_s=\sqrt{\nu^* T^*}$) at high frequency (refer to \S\ref{sec:gamma-small} when $\gamma \ll 1$), and the Ekman layer of the von K\'arm\'an viscous pump (of thickness $\delta^*_e=\sqrt{\nu^* D^*/(2W^*)}$) at high periods (refer to \S\ref{sec:gamma-large} when $\gamma \gg 1$).  It is therefore reasonable to expect that the wall turbulence over the disc surface is modified similarly to the oscillating-wall case at high frequency and to the steady-rotation case studied by RH13 at high periods. The parameter $\gamma$, written as $\gamma=(2/\pi)\left(\delta^*_s/\delta^*_e\right)^2$, can be interpreted as the threshold that distinguishes these two limiting regimes. The thinner boundary layer between these two limits 
dictates the way the turbulence is altered.  When $\gamma=\mathcal{O}(1)$, an intermediate oscillating-disc forcing regime is identified, for which viscous effects diffuse from the wall due to both unsteady oscillatory effects and to large-scale rotational motion. 

When $\gamma \ll 1$, the drag-reduction mechanism is analogous to the one advanced by \cite{ricco-etal-2012} for the oscillating-wall flow, namely that the near-wall periodic shear acts to increase the turbulent enstrophy and to attenuate the Reynolds stresses. 
Important differences from the oscillating-wall case are 
i) the wallward motion of high-speed fluid, entrained by the disc oscillation from the interior of the channel, 
ii) the radial-flow effects due to centrifugal forces, which are proportional to the nonlinear term $F'^2$ (refer to \eqref{eq:rosenblat-equations} for the laminar case) and produce additional spanwise forcing in planes perpendicular to the streamwise direction, 
iii) the radial dependence of the forcing amplitude, and iv) the degrading effect on drag reduction due to wall oscillations which are not spanwise oriented.
The latter effect was first documented by \citet{zhou-ball-2006}, who proved that spanwise wall oscillations produce the largest drag reduction, while streamwise wall oscillations lead to approximately a third of the spanwise-oscillation value. The shading on the disc surface in figure \ref{schematic} illustrates the effectiveness of the wall oscillations at different orientation angles. 
	
\item  {\em Role of quasi-steady inter-disc structures}

The second contribution is from the tubular interdisc structures, which are streamwise-elongated and quasi-steady as they persist throughout the disc oscillation. They are primarily synthetic jets, an indirect byproduct of the disc rotation (as in RH13) or disc oscillation. As discussed in \S\ref{sec:turbulent-discflow}, these jets are directed wallward and backward with respect to the mean flow $u_m$. The time-averaged flow between discs is therefore retarded with respect to the mean flow.  
Further insight into the generation of these structures could lead to other actuation methods leading to a similar drag reduction benefit. Although the structures appear directly above the regions of high shear created by neighbouring discs in the spanwise direction, they are largely unaffected by the time-modulation of the shear. These structures could be a product of the interaction between the radial streaming flows of neighbouring discs, which have a non-zero mean (refer to figure \ref{G-profile} (left)).
\end{itemize}
The FIK identity is useful because the role of disc boundary layer on drag reduction is distilled into $\mathcal{R}_t$, which sums up the decrease of turbulent Reynolds stresses, while the role of the structures is given by $\mathcal{R}_d$, which is solely due to the additional disc-flow Reynolds stresses. $\mathcal{R}_t$ and $\mathcal{R}_d$ quantify mathematically the two drag-reduction effects.   

It has been shown that drag reduction scales linearly with the penetration depth of the laminar layer for different spanwise wall forcing conditions, such as spatially uniform spanwise oscillation, travelling and steady wall waves \citep{ricco-etal-2012,cimarelli-etal-2013}. An analogous scaling is obtained in the following. The definition of the oscillating-wall penetration depth advanced by \cite{choi-xu-sung-2002} is modified to account for the viscous diffusion effects induced by the disc oscillation. \cite{choi-xu-sung-2002}'s definition is employed because it takes into account the influence of the wall forcing amplitude, which was not necessary in \cite{quadrio-ricco-2011} because the wave amplitude was constant. Following the discussion on the role of the disc boundary layer on drag reduction, the crucial point is that only $\mathcal{R}_t$, i.e. the portion of drag reduction related to the attenuation of the turbulent Reynolds stresses, is scaled with the penetration thickness. The scaling is carried 
out for the case with the largest diameter, $D=6.76$, for which the infinite-disc laminar flow solution best represents the disc boundary layer flow because of the limited interference between discs.

\begin{figure}
\centering
\includegraphics[width=\textwidth]{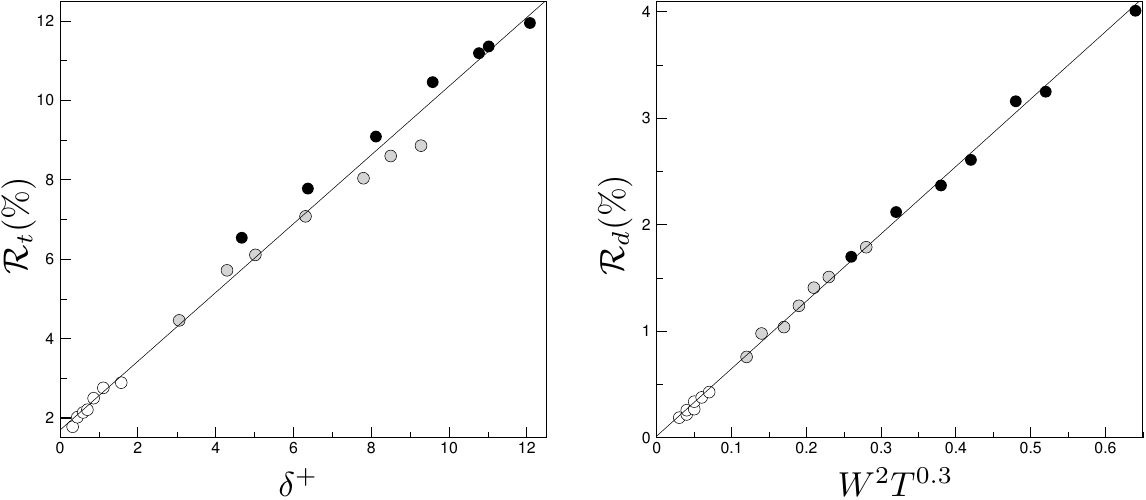}
\vspace{0.0cm}
\caption{
Left: $\mathcal{R}_t$, the contribution to drag reduction due to turbulent Reynolds stress attenuation, vs. $\delta^+$, the penetration depth, defined in \eqref{eq:delta}. Right: $\mathcal{R}_d$, the contribution to drag reduction due to the disc-flow Reynolds stresses, vs. $W^2T^{0.3}$. The diameter is $D=6.76$. White circles: $W^+=3$, light grey: $W^+=6$, black: $W^+=9$. } 
\label{scaling}
\end{figure}

From the envelope of the Stokes layer velocity profile engendered by an oscillating wall
\begin{equation*}
W_e^+=W_m^+\exp\left( -\sqrt{\pi/T^+}y^+\right)\text{,}
\end{equation*}
\citet{choi-xu-sung-2002} defined the penetration depth as
\begin{equation*}
y_d^+=\sqrt{T^+/\pi}\ln\left(W_m^+/W_{th}^+\right)\text{,}
\end{equation*}
where $W_m^+$ is the maximum wall velocity and $W_{th}^+$ is a threshold value below which the induced spanwise oscillations have little effect on the channel flow. For the oscillating disc case, the enveloping function for the laminar azimuthal disc velocity, $W_e^+=W^+G_e(\eta,\gamma)$, where
\begin{equation*}
G_e(\eta,\gamma)=\max_{\breve{t}}G(\eta,\breve{t},\gamma)\text{,}
\end{equation*}
plays a role analogous to the exponential envelope for the classical Stokes layer. Defining the inverse of $G_e$, $\mathsf{L}=G_e^{-1}$, the penetration depth of the oscillating-disc layer is obtained as
\begin{equation}
\label{eq:delta}
\delta^+=\sqrt{T^+/\pi}\hspace{1mm}\mathsf{L}\left(W^+/W_{th}^+\right)\text{.}
\end{equation}
Note that in the limit of $\gamma\rightarrow0$ one finds
\begin{equation*}
\lim_{\gamma\rightarrow0}\mathsf{L}\left(W^+/W_{th}^+\right)=\ln\left(W^+/W_{th}^+\right)\text{.}
\end{equation*}
The Stokes layer penetration depth is therefore obtained as a special case. In figure \ref{scaling} (left), the drag-reduction contributor $\mathcal{R}_t$ shows a satisfactory linear scaling with the penetration depth, computed via \eqref{eq:delta} with $W^+_{th}=2.25$.

In order to find a scaling for $\mathcal{R}_d$, the portion of drag reduction only due to the inter-disc structures, the FIK identity and the laminar solution discussed in \S\ref{sec:laminar} are employed. From \eqref{rd}, it is evident that $\mathcal{R}_d$ is proportional to $\widehat{u_dv_d}$. Through the definitions of the laminar velocity components \eqref{velocity-relations}, $u_d\sim W$ and $v_d\sim W\sqrt{T}$. It then follows that a reasonable estimate could be $u_dv_d\sim W^2\sqrt{T}$ at the edge of the discs where the structures appear. It is then logical to look for a scaling of $\mathcal{R}_d$ in the form $W^mT^n$. An excellent linear fit for the drag reduction data is found for $(m,n)=(2,0.3$), 
as shown in figure \ref{scaling} (right). Outer-unit scaling for $W$ and $T$ applies, which means that the structures are not influenced by the change in $u^*_{\tau}$. The exponent of $W$ is as predicted by the laminar solution. The deviation of the coefficient $n$ from that predicted by the laminar analysis (i.e. $n=0.5$) can be accounted for by the factors which are not taken into account in the laminar analysis, such as the disc-flow interaction with the streamwise turbulent flow and between neighbouring discs. 

\section{Outlook for the future}
\label{sec:outlook}

In line with the analysis by RH13 for the steady disc-flow technique, it is instructive to render the scaled disc forcing parameters dimensional to guide laboratory experiments and to estimate the characteristic length and time scales of the wall forcing for flows of technological relevance. Table \ref{tab:outlook} displays estimated data for three flows of industrial interest and two flows of experimental interest with $D=6.76$, $W=0.39$, and $T=130$, which lead to $\mathcal{R}=16\%$ and $\mathcal{P}_{net}=5.5\%$. 
This table may be compared with the analogous table 6 in RH13 for the steady rotation case, although it should be noted that $f^*$ indicates the oscillation frequency in the present case ($f^*=2\pi/T^*$) and the rotational frequency in RH13's case ($f^*=\omega^*/2\pi$, where 
$\omega^*$ is the angular velocity).

Experimental realisation of the disc-flow technique is possible with $D^*=4-8\hspace{1mm}\mbox{cm}$, $W^*=0.2\hspace{1mm}\mbox{m/s}$ in a water channel and $4.6\hspace{1mm}\mbox{m/s}$ in a wind tunnel. The frequencies are $f^*=0.37\hspace{1mm}\mbox{Hz}$ and $16\hspace{1mm}\mbox{Hz}$, respectively. The dimensional parameters in flight are $D^*=5.8\hspace{1mm}\mbox{mm}$, $W^*=70.7\hspace{1mm}\mbox{m/s}$, and $f^*=1752\hspace{1mm}\mbox{Hz}$. Commercially available electromagnetic motors ($D^*=2\hspace{1mm}\mbox{mm}$, $f^*=\mathcal{O}(10^{3})\hspace{1mm}\mbox{Hz}$), adapted for oscillatory motion, would guarantee these time and length scales of forcing \citep{chenliu-etal-2010}. The optimal frequency in flight is approximately half of the optimal one for steady rotation: $f^*=1752\hspace{1mm}\mbox{Hz}$ for the oscillating discs compared to $f^*=3718\hspace{1mm}\mbox{Hz}$ for the steady rotating discs. 

\begin{figure}
\centering
\includegraphics[width=0.8\textwidth]{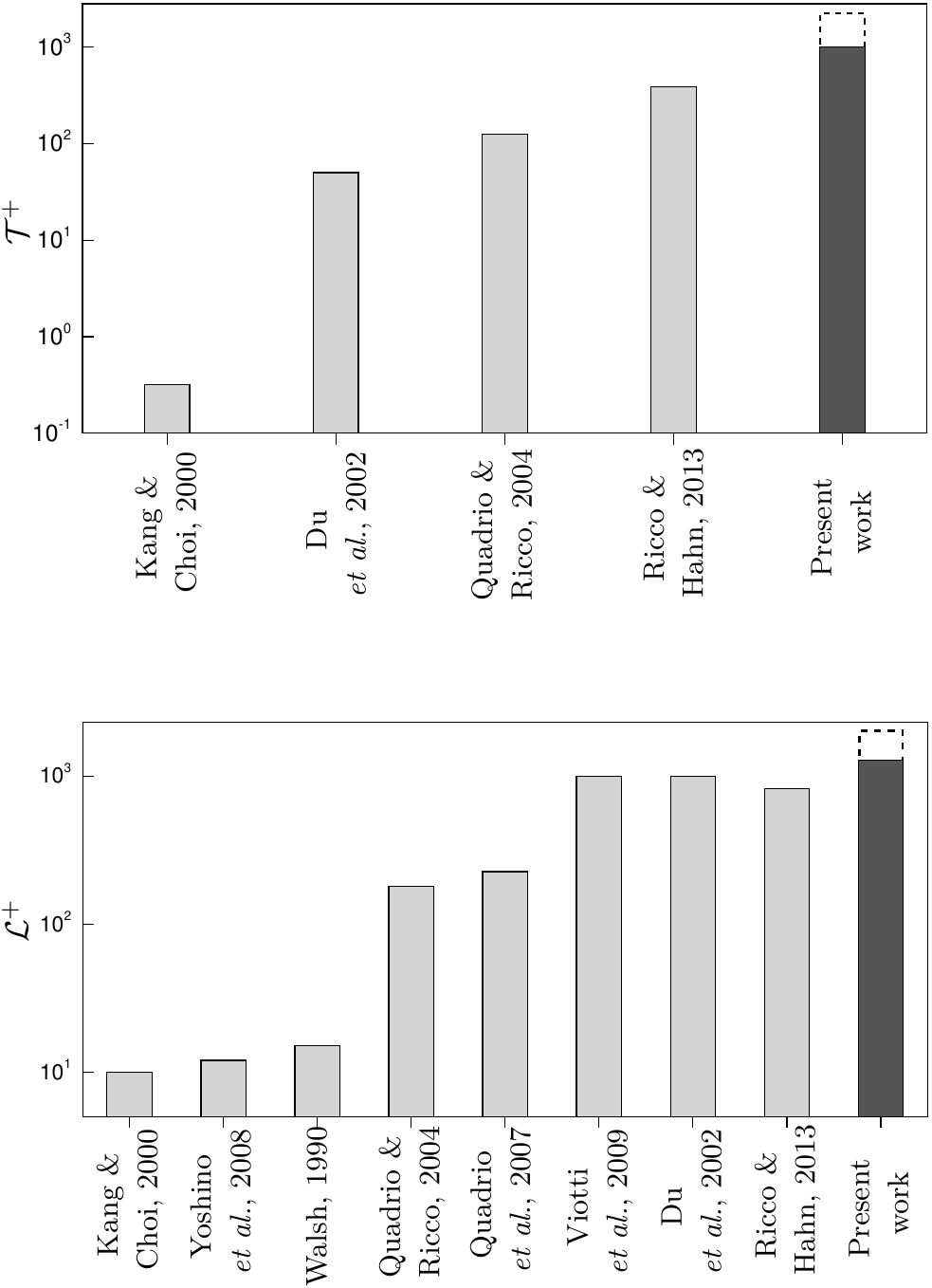}
\caption{Characteristic optimal time and length scales, $\mathcal{T}^{+}$, $\mathcal{L}^{+}$, for a range of drag reduction methods are shown for comparison with the oscillating disc technique.  From left to right the time scales are given as follows:  time between successive flow field measurements \citep{kang-choi-2000}, period of transverse travelling wave forcing \citep{du-symeonidis-karniadakis-2002}, period of spanwise wall oscillations \citep{quadrio-ricco-2004}, period of rotation of steady disc forcing \citep{ricco-hahn-2013}, and period of disc oscillation.  From left to right the length scales are given as follows: maximum displacement of wall-normal wall motions \citep{kang-choi-2000}, spacing of sensors for feedback control of wall deformation \citep{yoshino-suzuki-kasagi-2008}, riblet spacing \citep{walsh-1990}, maximum displacement of temporally oscillating wall \citep{quadrio-ricco-2004}, wavelength of streamwise-sinusoidal wall transpiration \citep{quadrio-floryan-luchini-2007}, wavelength 
of standing wave forcing \citep{viotti-quadrio-luchini-2009}, wavelength of transverse travelling wave forcing \citep{du-symeonidis-karniadakis-2002}, diameter of steady discs \citep{ricco-hahn-2013}, and diameter of oscillating discs.}
\label{scales}
\end{figure}

Figure~\ref{scales} shows characteristic time and length scales of the oscillating-disc technique and of other drag reduction methods. 
The typical length scale of the oscillating-disc technique is larger than that of the steadily rotating discs and the standing wave forcing, whilst being two orders of magnitude greater than both riblets and the feedback control systems studied by \citet{yoshino-suzuki-kasagi-2008}. The typical time scale of the oscillating disc flow is one order of magnitude larger than that of the oscillating wall forcing.
It is also worth pointing out that these are optimal values for the tested parameter range and that our results in \S\ref{sec:turbulent-dependence} hint at the possibility to obtain comparable drag-reduction values for even larger oscillation periods and diameters, which are 
denoted by the dashed lines in figure \ref{scales}.

The notable limitation of our analysis is the low Reynolds number of the simulations. It is therefore paramount to investigate the disc-flow properties at higher Reynolds number to assess whether and how the maximum drag reduction values and the optimal forcing conditions vary.  

We close our discussion by mentioning another advantage of the oscillating-disc flow when compared to the steady-disc flow by RH13. As shown in figure \ref{D-hr-outer} (d), it is possible to achieve $\mathcal{R}=13\%$ with $\gamma=\pi/8$, $T=12$, $W=0.51$, i.e. the disc tip undertakes a maximum displacement of only $1/8$ of the disc circumference. Therefore, for this case the disc-flow technique could be realized in a laboratory by use of a thin elastic seal between the disc and the stationary wall. This design would eliminate any clearance around the discs, which would not be possible for the case of steady rotation.   

\renewcommand{\arraystretch}{1.2}
\begin{table}
\centering
\begin{tabular}{l||c|c|c||c|c}
Parameter & Flight (BL) & Ship (BL) & Train (BL) & WT (BL) & WC (CF) \\
\hline \hline
$U^*\hspace{2mm}\mbox{(m/s)}$ & $225$ & $10$ & $83$ & $11.6$ & $0.4$ \\
$\nu^*\cdot10^6\hspace{2mm}\mbox{(m$^{2}$/s)}$ & $35.3$ & $1.5$ & $15.7$ & $15.7$ & $1.1$ \\
$x^*\hspace{2mm}\mbox{(m)}$ & $1.5$ & $1.5$ & $1.8$ & $1.0$ & - \\
$h^*\hspace{2mm}\mbox{(mm)}$ & $22$ & $22$ & $27$ & $25$ & $10$ \\
$u_\tau^*\hspace{2mm}\mbox{(m/s)}$ & $7.9$ & $0.4$ & $2.9$ & $0.5$ & $0.02$ \\
$Re_\tau$ & $4970$ & $4970$ & $4970$ & $800$ & $180$ \\
$C_{f}\cdot10^{3}$ & $2.4$ & $2.4$ & $2.4$ & $3.8$ & $8.1$ \\
\hline
$D^*\hspace{2mm}\mbox{(mm)}$ & $5.7$ & $5.6$ & $6.9$ & $39.6$ & $70.9$ \\
$W^*\hspace{2mm}\mbox{(m/s)}$ & $70.7$ & $3.1$ & $26.1$ & $4.6$ & $0.2$ \\
$T^*\hspace{2mm}\mbox{(ms)}$ & $0.6$ & $12.5$ & $1.9$ & $61$ & $2700$ \\
$f^*\hspace{2mm}\mbox{(Hz)}$ & $1752$ & $80$ & $536$ & $16$ & $0.4$ 
\end{tabular}
\caption{Dimensional quantities for the optimum $\mathcal{P}_{net}$ case for three flows of industrial and two of experimental interest ($D=6.76$, $W=0.39$ and $T=130$). In the headings (BL) indicates a turbulent boundary layer with no pressure gradient, and (CF) indicates a pressure-driven channel flow.  WT and WC stand for wind tunnel and water channel respectively.  For headings marked BL, $U^*$ represents the free-stream mean velocity, $x^*$ is the downstream location and $h^*$ the boundary layer thickness; whilst for the CF case $U^*$ represents the bulk velocity and $h^*$ the channel half-height.  The relations used: $h^*=0.37x^*(x^*U^*/\nu^*)^{-0.2}$ and $C_{f}=0.37\left[ \log_{10}(x^*U^*/\nu^*) \right]^{-2.584}$ for BL; $C_{f}=0.0336Re_\tau^{-0.273}$ for CF are from \citet{pope-2000}.}
\label{tab:outlook}
\end{table}

\nocite{kang-choi-2000,du-symeonidis-karniadakis-2002,quadrio-floryan-luchini-2007,walsh-1990}

\section*{Acknowledgements}
We would like to thank the Department of Mechanical Engineering at the University of Sheffield for funding this research. This work would have not been possible without the use of the computing facilities of N8 HPC, funded by the N8 consortium and EPSRC (Grant EP/K000225/1). The Centre is coordinated by the Universities of Leeds and Manchester. We also acknowledge the help of Dr Chris Davies from the University of Cardiff and Mr Harry Day from the University of Sheffield on the numerical computation of the laminar oscillating-disc flow. We are also indebted to Professors Shuisheng He, Ning Qin, and Yang Zhang, Dr Bryn Jones, and Misses Elena Marensi and Claudia Alvarenga at the University of Sheffield, Dr Ati Sharma at the University of Southampton, and Dr Hasegawa at the University of Tokyo for providing insightful comments on a preliminary version of the manuscript. Part of this work was presented at the 66$^{\mbox{th}}$ Annual Meeting of the APS Division of Fluid Dynamics, Pittsburgh, 
Pennsylvania, in November 2013.

\bibliographystyle{jfm}
\bibliography{pr}

\begin{thebibliography}{49}
\expandafter\ifx\csname natexlab\endcsname\relax\def\natexlab#1{#1}\fi

\bibitem[Abramowitz \& Stegun(1964)]{abramowitz-stegun-1964}
{\sc Abramowitz, M. \& Stegun, I.A.} 1964 {\em {H}andbook of {M}athematical
  {F}unctions\/}. Nat. Bureau Stand. Appl. Math. Ser. 55.

\bibitem[Bandyopadhyay(2006)]{bandyopadhyay-2006}
{\sc Bandyopadhyay, P.~R.} 2006 Stokes mechanism of drag reduction. {\em J.
  Appl. Mech.\/} {\bf 73}, 483--489.

\bibitem[Baron \& Quadrio(1996)]{baron-quadrio-1996}
{\sc Baron, A. \& Quadrio, M.} 1996 Turbulent drag reduction by spanwise wall
  oscillations. {\em Appl. Sc. Res.\/} {\bf 55}, 311--326.

\bibitem[Batchelor(1967)]{batchelor-1967}
{\sc Batchelor, G.~K.} 1967 {\em {A}n {I}ntroduction to {F}luid {D}ynamics\/}.
  Cambridge University Press.

\bibitem[Benney(1964)]{benney-1964}
{\sc Benney, D.J.} 1964 The flow induced by a disk oscillating in its own
  plane. {\em J. Fluid Mech.\/} {\bf 18}~(03), 385--391.

\bibitem[Berger {\em et~al.\/}(2000)Berger, Kim, Lee \& Lim]{berger-etal-2000}
{\sc Berger, T.W., Kim, J., Lee, C. \& Lim, J.} 2000 Turbulent boundary layer
  control utilizing the {L}orentz force. {\em Phys. Fluids\/} {\bf 12}~(3),
  631--649.

\bibitem[Canuto {\em et~al.\/}(2007)Canuto, Hussaini, Quarteroni \&
  Zang]{canuto-etal-2007}
{\sc Canuto, C., Hussaini, M.Y., Quarteroni, A. \& Zang, T.A.} 2007 {\em
  Spectral methods: evolution to complex geometries and applications to fluid
  dynamics\/}. Springer.

\bibitem[Carmi \& Tustaniwskyj(1981)]{carmi-tustaniwskyj-1981}
{\sc Carmi, S. \& Tustaniwskyj, J.I.} 1981 Stability of modulated finite-gap
  cylindrical {C}ouette flow: linear theory. {\em J. Fluid Mech.\/} {\bf 108},
  19--42.

\bibitem[Choi {\em et~al.\/}(2002)Choi, Xu \& Sung]{choi-xu-sung-2002}
{\sc Choi, J-I., Xu, C-X. \& Sung, H.J.} 2002 Drag reduction by spanwise wall
  oscillation in wall-bounded turbulent flows. {\em AIAA J.\/} {\bf 40}~(5),
  842--850.

\bibitem[Cimarelli {\em et~al.\/}(2013)Cimarelli, Frohnapfel, Hasegawa,
  De~Angelis \& Quadrio]{cimarelli-etal-2013}
{\sc Cimarelli, A., Frohnapfel, B., Hasegawa, Y., De~Angelis, E. \& Quadrio,
  M.} 2013 Prediction of turbulence control for arbitrary periodic spanwise
  wall movement. {\em Phys. Fluids\/} {\bf 25}~(075102).

\bibitem[Dhanak \& Si(1999)]{dhanak-si-1999}
{\sc Dhanak, M.R. \& Si, C.} 1999 On reduction of turbulent wall friction
  through spanwise oscillations. {\em J. Fluid Mech.\/} {\bf 383}, 175--195.

\bibitem[Di~Cicca {\em et~al.\/}(2002)Di~Cicca, Iuso, Spazzini \&
  Onorato]{dicicca-etal-2002}
{\sc Di~Cicca, G.~M., Iuso, G., Spazzini, P.~G \& Onorato, M.} 2002 Particle
  image velocimetry investigation of a turbulent boundary layer manipulated by
  spanwise oscillations. {\em J. Fluid Mech.\/} {\bf 467}, 41--56.

\bibitem[Du {\em et~al.\/}(2002)Du, Symeonidis \&
  Karniadakis]{du-symeonidis-karniadakis-2002}
{\sc Du, Y., Symeonidis, V. \& Karniadakis, G.E.} 2002 Drag reduction in
  wall-bounded turbulence via a transverse travelling wave. {\em J. Fluid
  Mech.\/} {\bf 457}, 1--4.

\bibitem[Duque-Daza {\em et~al.\/}(2012)Duque-Daza, Baig, Lockerby,
  Chernyshenko \& Davies]{duque-etal-2012}
{\sc Duque-Daza, C.A., Baig, M.F., Lockerby, D.A., Chernyshenko, S.I. \&
  Davies, C.} 2012 {M}odelling turbulent skin-friction control using linearized
  {N}avier-{S}tokes equations. {\em J. Fluid Mech.\/} {\bf 702}, 403--414.

\bibitem[Fukagata {\em et~al.\/}(2002)Fukagata, Iwamoto \&
  Kasagi]{fukagata-iwamoto-kasagi-2002}
{\sc Fukagata, K., Iwamoto, K. \& Kasagi, N.} 2002 Contribution of {R}eynolds
  stress distribution to the skin friction in wall-bounded flows. {\em Phys.
  Fluids\/} {\bf 14}~(11), 73--76.

\bibitem[Gibson(2006)]{gibson-2006}
{\sc Gibson, J.~F.} 2006 Channelflow: a spectral {N}avier-{S}tokes simulator in
  {C}++. ``\url{http://www.channelflow.org/}''.

\bibitem[Gouder {\em et~al.\/}(2013)Gouder, Potter \&
  Morrison]{gouder-potter-morrison-2013}
{\sc Gouder, K., Potter, M. \& Morrison, J.F.} 2013 Turbulent friction drag
  reduction using electroactive polymer and electromagnetically driven
  surfaces. {\em Exp. Fluids\/} {\bf 54}, 1--12.

\bibitem[Hinze(1975)]{hinze-1975}
{\sc Hinze, J.O.} 1975 {\em Turbulence\/}. McGraw Hill, Inc. -- Second Edition.

\bibitem[Iuso {\em et~al.\/}(2003)Iuso, Di~Cicca, Onorato, Spazzini \&
  R.]{iuso-etal-2003}
{\sc Iuso, G., Di~Cicca, G.~M., Onorato, M., Spazzini, P.~G. \& R., Malvano}
  2003 Velocity streak structure modifications induced by flow manipulation.
  {\em Phys. Fluids\/} {\bf 15}~(9), 2602--2612.

\bibitem[Jung {\em et~al.\/}(1992)Jung, Mangiavacchi \&
  Akhavan]{jung-mangiavacchi-akhavan-1992}
{\sc Jung, W.J., Mangiavacchi, N. \& Akhavan, R.} 1992 Suppression of
  turbulence in wall-bounded flows by high-frequency spanwise oscillations.
  {\em Phys. Fluids A\/} {\bf 4}~(8), 1605--1607.

\bibitem[Kang \& Choi(2000)]{kang-choi-2000}
{\sc Kang, S. \& Choi, H.} 2000 Active wall motions for skin-friction drag
  reduction. {\em Phys. Fluids\/} {\bf 12}~(12), 3301--3304.

\bibitem[Kasagi {\em et~al.\/}(2009)Kasagi, Suzuki \&
  Fukagata]{kasagi-suzuki-fukagata-2009}
{\sc Kasagi, N., Suzuki, Y. \& Fukagata, K.} 2009 Micromechanical systems-based
  feedback control of turbulence for skin friction reduction. {\em Ann. Rev.
  Fluid Mech.\/} {\bf 41}, 231--251.

\bibitem[Keefe(1998)]{keefe-1998}
{\sc Keefe, L.} 1998 Method and apparatus for reducing the drag of flows over
  surfaces. {\em United States Patent\/} {\bf 5,803,409}.

\bibitem[Kleiser \& Schumann(1980)]{kleiser-schumann-1980}
{\sc Kleiser, L. \& Schumann, U.} 1980 Treatment of incompressibility and
  boundary conditions in 3-{D} numerical spectral simulations of plane channel
  flows. In {\em Proc. 3rd {GAMM} {C}onf. {N}umerical {M}ethods in {F}luid
  {M}echanics\/} (ed. E.~Hirschel), pp. 165--173. GAMM, Vieweg.

\bibitem[Kuang-Chen~Liu {\em et~al.\/}(2010)Kuang-Chen~Liu, Friend \&
  Yeo]{chenliu-etal-2010}
{\sc Kuang-Chen~Liu, D., Friend, J. \& Yeo, L.} 2010 A brief review of
  actuation at the micro-scale using electrostatics, electromagnetics and
  piezoelectric ultrasonics. {\em Acoust. Sci. \& Tech.\/} {\bf 31}, 115 --
  123.

\bibitem[Laadhari {\em et~al.\/}(1994)Laadhari, Skandaji \&
  Morel]{laadhari-skandaji-morel-1994}
{\sc Laadhari, F., Skandaji, L. \& Morel, R.} 1994 Turbulence reduction in a
  boundary layer by local spanwise oscillating surface. {\em Phys. Fluids\/}
  {\bf 6}~(10), 3218--3220.

\bibitem[Moarref \& Jovanovic(2012)]{moarref-jovanovic-2012}
{\sc Moarref, R. \& Jovanovic, M.R.} 2012 Model-based design of transverse wall
  oscillations for turbulent drag reduction. {\em J. Fluid Mech.\/} {\bf 707},
  205--240.

\bibitem[Panton(1995)]{panton-1995}
{\sc Panton, R.} 1995 {\em Incompressible {F}low\/}. Wiley-Interscience --
  Second Edition.

\bibitem[Pope(2000)]{pope-2000}
{\sc Pope, S.B.} 2000 {\em Turbulent {F}lows\/}. Cambridge University Press.

\bibitem[Pozrikidis(2009)]{pozrikidis-2009}
{\sc Pozrikidis, C.} 2009 {\em Fluid Dynamics: Theory, Computation, and
  Numerical Simulation\/}. Springer.

\bibitem[Quadrio(2011)]{quadrio-2011}
{\sc Quadrio, M.} 2011 Drag reduction in turbulent boundary layers by in-plane
  wall motion. {\em Phil. Trans. Royal Soc. A\/} {\bf 369}~(1940), 1428--1442.

\bibitem[Quadrio {\em et~al.\/}(2007)Quadrio, Floryan \&
  Luchini]{quadrio-floryan-luchini-2007}
{\sc Quadrio, M., Floryan, J.M. \& Luchini, P.} 2007 Effect of
  streamwise-periodic wall transpiration on turbulent friction drag. {\em J.
  Fluid Mech.\/} {\bf 576}, 424--444.

\bibitem[Quadrio \& Ricco(2004)]{quadrio-ricco-2004}
{\sc Quadrio, M. \& Ricco, P.} 2004 Critical assessment of turbulent drag
  reduction through spanwise wall oscillations. {\em J. Fluid Mech.\/} {\bf
  521}, 251--271.

\bibitem[Quadrio \& Ricco(2011)]{quadrio-ricco-2011}
{\sc Quadrio, M. \& Ricco, P.} 2011 The laminar generalized {S}tokes layer and
  turbulent drag reduction. {\em J. Fluid Mech.\/} {\bf 667}, 135--157.

\bibitem[Quadrio {\em et~al.\/}(2009)Quadrio, Ricco \&
  Viotti]{quadrio-ricco-viotti-2009}
{\sc Quadrio, M., Ricco, P. \& Viotti, C.} 2009 Streamwise-travelling waves of
  spanwise wall velocity for turbulent drag reduction. {\em J. Fluid Mech.\/}
  {\bf 627}, 161--178.

\bibitem[Quadrio \& Sibilla(2000)]{quadrio-sibilla-2000}
{\sc Quadrio, M. \& Sibilla, S.} 2000 Numerical simulation of turbulent flow in
  a pipe oscillating around its axis. {\em J. Fluid Mech.\/} {\bf 424},
  217--241.

\bibitem[Ricco \& Hahn(2013)]{ricco-hahn-2013}
{\sc Ricco, P. \& Hahn, S.} 2013 Turbulent drag reduction through rotating
  discs. {\em J. Fluid Mech.\/} {\bf 722}, 267--290.

\bibitem[Ricco {\em et~al.\/}(2012)Ricco, Ottonelli, Hasegawa \&
  Quadrio]{ricco-etal-2012}
{\sc Ricco, P., Ottonelli, C., Hasegawa, Y. \& Quadrio, M.} 2012 Changes in
  turbulent dissipation in a channel flow with oscillating walls. {\em J. Fluid
  Mech.\/} {\bf 700}, 77--104.

\bibitem[Ricco \& Quadrio(2008)]{ricco-quadrio-2008}
{\sc Ricco, P. \& Quadrio, M.} 2008 Wall-oscillation conditions for drag
  reduction in turbulent channel flow. {\em Int. J. Heat Fluid Flow\/} {\bf
  29}, 601--612.

\bibitem[Rogers \& Lance(1960)]{rogers-lance-1960}
{\sc Rogers, M.H. \& Lance, G.N.} 1960 The rotationally symmetric flow of a
  viscous fluid in the presence of an infinite rotating disk. {\em J. Fluid
  Mech.\/} {\bf 7}~(4), 617--631.

\bibitem[Rosenblat(1959)]{rosenblat-1959}
{\sc Rosenblat, S.} 1959 Torsional oscillations of a plane in a viscous fluid.
  {\em J. Fluid Mech.\/} {\bf 6}~(2), 206--220.

\bibitem[Skote(2011)]{skote-2011}
{\sc Skote, M.} 2011 Turbulent boundary layer flow subject to streamwise
  oscillation of spanwise wall-velocity. {\em Phys. Fluids\/} {\bf 23}, 081703.

\bibitem[Skote(2013)]{skote-2013}
{\sc Skote, M.} 2013 Comparison between spatial and temporal wall oscillations
  in turbulent boundary layer flows. {\em J. Fluid Mech.\/} {\bf 730},
  273--294.

\bibitem[Trujillo {\em et~al.\/}(1997)Trujillo, Bogard \&
  Ball]{trujillo-bogard-ball-1997}
{\sc Trujillo, S.M., Bogard, D.G. \& Ball, K.S.} 1997 Turbulent boundary layer
  drag reduction using an oscillating wall. {\em AIAA Paper\/} {\bf 97-1870}.

\bibitem[Viotti {\em et~al.\/}(2009)Viotti, Quadrio \&
  Luchini]{viotti-quadrio-luchini-2009}
{\sc Viotti, C., Quadrio, M. \& Luchini, P.} 2009 Streamwise oscillation of
  spanwise velocity at the wall of a channel for turbulent drag reduction. {\em
  Phys. Fluids\/} {\bf 21}~(115109).

\bibitem[Walsh(1990)]{walsh-1990}
{\sc Walsh, M.~J.} 1990 Riblets. In {\em Viscous {D}rag {R}eduction in
  {B}oundary {L}ayers\/} (ed. D.~M. Bushnell \& J.~N. Hefner), , vol. 123, pp.
  203--261. Progress in Astronautics and Aeronautics.

\bibitem[Wang(1989)]{wang-1989}
{\sc Wang, C.Y.} 1989 Shear flow over a rotating plate. {\em App. Sc. Res.\/}
  {\bf 46}, 89--96.

\bibitem[Yoshino {\em et~al.\/}(2008)Yoshino, Suzuki \&
  Kasagi]{yoshino-suzuki-kasagi-2008}
{\sc Yoshino, T., Suzuki, Y. \& Kasagi, N.} 2008 Drag reduction of turbulence
  air channel flow with distributed micro-sensors and actuators. {\em J. Fluid
  Sci. Technol.\/} {\bf 3}, 137--148.

\bibitem[Zhou \& Ball(2008)]{zhou-ball-2006}
{\sc Zhou, D. \& Ball, K.S.} 2008 Turbulent drag reduction by spanwise wall
  oscillations. {\em Int. J. Eng. Trans. A Basics\/} {\bf 21}~(1), 85.

\end{thebibliography}

\end{document}